\documentclass[10pt]{article}

\usepackage{afterpage}
\usepackage{float}
\usepackage{longtable}
\usepackage{graphicx}
\usepackage{pdflscape}
\usepackage[numbers,sort&compress]{natbib}
\usepackage{psfrag}
\usepackage[unicode=true,
  linktocpage,
  linkbordercolor={0.5 0.5 1},
  citebordercolor={0.5 1 0.5},
  linkcolor=blue]{hyperref}

\usepackage{enumitem}
\usepackage{amssymb}
\usepackage{amsmath}
\usepackage{amssymb}
\usepackage{multirow}
\usepackage{booktabs}
\usepackage[]{algorithm2e}
\usepackage{float}
\usepackage{array}\newcolumntype{L}[1]{>{\raggedright\let\newline\\\arraybackslash\hspace{0pt}}m{#1}}
\newcolumntype{C}[1]{>{\centering\let\newline\\\arraybackslash\hspace{0pt}}m{#1}}
\newcolumntype{R}[1]{>{\raggedleft\let\newline\\\arraybackslash\hspace{0pt}}m{#1}}

\usepackage{graphicx}
\usepackage{amssymb}
\usepackage{amsmath}
\usepackage{multirow}
\usepackage{booktabs}
\usepackage{array}
\usepackage{todonotes}
\newcolumntype{L}[1]{>{\raggedright\let\newline\\\arraybackslash\hspace{0pt}}m{#1}}
\newcolumntype{C}[1]{>{\centering\let\newline\\\arraybackslash\hspace{0pt}}m{#1}}
\newcolumntype{R}[1]{>{\raggedleft\let\newline\\\arraybackslash\hspace{0pt}}m{#1}}

\title{Sparse and redundant signal representations for x-ray computed tomography}

\author{Davood Karimi}

\begin{document}

\maketitle

\section{A brief history of x-ray computed tomography}

Computed tomography (CT) refers to creating images of the cross sections of an object using transmission or reflection data. These data are usually referred to as the \textit{projections} of the object. For the projection data to be sufficient, the object needs to be illuminated from many different directions. The problem of reconstructing the image of an object from its projections has various applications, from  reconstructing the structure of molecules from data collected with electron microscopes to reconstructing maps of radio emissions of celestial objects from data collected with radio telescopes \cite{herman2009}. However, the most important applications of CT have been in the field of medicine, where the impact of CT has been nothing short of revolutionary. Today, physicians and surgeons are able to view the internal organs of their patients with a precision and safety that was impossible to imagine before the advent of CT.

Most of the medical imaging modalities including ultrasound, magnetic resonance imaging (MRI), and positron emission tomography (PET) can be considered as examples of CT. The fundamental difference between these modalities is the property of the material (i.e., tissue) that they image. X-ray CT, which is our focusn, is based on the tissue's ability to attenuate x-ray photons. X-rays had been discovered by the German physicist Wilhelm Rontgen in 1895. Rontgen, who won the first Nobel Prize in Physics for this discovery, realized that x-rays could reveal the skeletal structure of the body parts because bones and soft tissue had different x-ray attenuation properties. However, the first commercial CT scanners appeared in the early 1970s, finally winning the 1979 Nobel Prize in Medicine for Allan Cormack and Godfrey Hounsfield for independently inventing CT.

Today, x-ray CT is an indispensable tool in medicine. In fact, the words CT and computed tomography are used to refer to x-ray CT with no confusion. Since its commercial introduction more than 40 years ago, diagnostic and therapeutic applications of CT have continued to grow. In the past two decades, especially, great advancements have been made in CT scanner technology and the available computational resources. Moreover, new scanning methods such as dual-source and dual-energy CT have become commercially available. Today, very fast scanning of large volumes has become possible. This has led to a dramatic increase in CT usage in clinical settings. It is estimated that globally more than 50,000 dual-energy x-ray CT scanners are in operation \cite{shepherd2014}. In the USA alone, the number of CT scans made annually increased from 19 million to 62 million between 1993 and 2006 \cite{mettler2008}.

\section{Imaging model}
\label{sec:CT_Imaging_Model}

\begin{figure}[ht]
    \centering
    \includegraphics[width=0.7\textwidth]{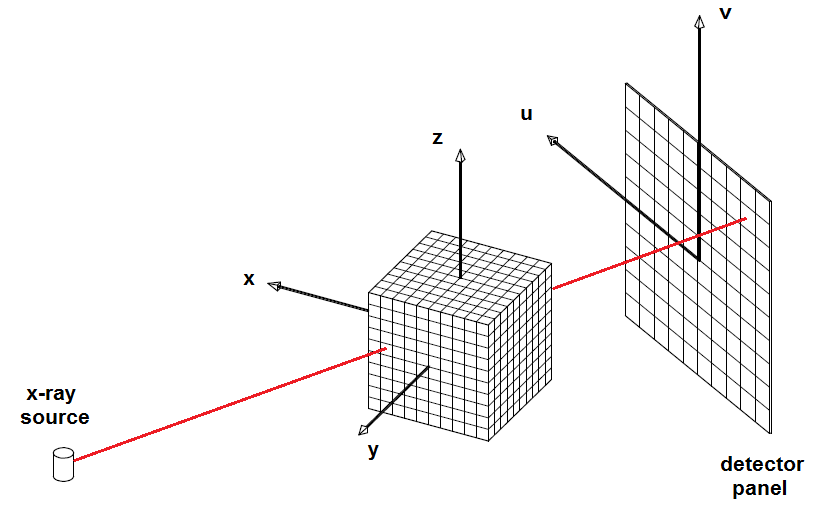}
    \caption{A schematic representation of cone-beam CT geometry.}
    \label{fig:cbct_geometry}
\end{figure}

Cone-beam computed tomography (CBCT) is a relatively new scan geometry that has found applications as diverse as image-guided radiation therapy, dentistry, breast CT, and microtomography \cite{scarfe2008, chen2002, ford2003, jaffray2002}. Figure \ref{fig:cbct_geometry} shows a schematic representation of CBCT. Divergent x-rays penetrate the object and become attenuated before being detected by an array of detectors. The equation relating the detected photon number to the line integral of the attenuation coefficient is \cite{vanmatter2000}:

\begin{equation} \label{eq:modelc}
\frac{N_d^i}{N_0^i}= \exp \left(-\int_i \mu ds \right)
\end{equation}

\noindent where $N_0^i$ and $N_d^i$ denote, respectively, the emitted and detected photon numbers for the ray from the x-ray source to the detector bin $i$ and $\int_i \mu ds$ is the line integral of the attenuation coefficient along that ray. By discretizing the imaged object, the following approximation to \eqref{eq:modelc} can be made:

\begin{equation} \label{eq:modeld}
\log\left(\frac{N_0^i}{N_d^i}\right)= \sum_{k=1}^{K} w_k^i x_k 
\end{equation}

\noindent where $x_k$ is the value of the unknown image at voxel $k$ and $w_k^i$ is the length of intersection of ray $i$ with this voxel. The equations for all measurements can be combined and conveniently written in matrix form as:

\begin{equation} \label{eq:modelm}
y= Ax + w
\end{equation}

\noindent where $y$ represents the vector of measurements (also known as the sinogram), $x$ is the unknown image, $A$ represents the projection matrix, and $w$ is the measurement noise.

The discretization approach mentioned above has several shortcomings. For example, it does not consider the finite size of the x-ray source and the detector area. Furthermore, exact computation of the intersection lengths of rays with voxels is computationally very costly for large-scale 3D CT. Therefore, several efficient implementations of the system matrix $A$ have been proposed \cite{long2010, lewitt1992, matej1996, deman2004}. For large-scale 3D CT, matrix $A$ it too large to be saved in computer memory. Instead, these algorithms implement multiplication with matrix $A$ and its transpose by computing the matrix elements on-the-fly.

Even though in theory $N_d$ follows a Poisson distribution, due to many complicating factors including the polychromatic nature of the x-ray source and the electronic noise, an accurate model of the raw data takes the form of a compound Poisson, shifted Poisson, or Poisson+Gaussian distribution \cite{lariviere2005}. For many practical applications, an adequate noise model is obtained by adding a Gaussian noise (to simulate the electronic noise) to the theoretical values of $N_d$. More realistic modeling, especially in low-dose CT, is much more complex and will need to take into account very subtle phenomena, which are the subject of much research \cite{thibault2006, nuyts2013, zabic2013}. An alternative approach is to consider the ratio of the photon counts after the logarithm transformation. Even though $N_0^i$ and $N_d^i$ are Poisson distributed, the noise in the sinogram (i.e., after the logarithm transformation) can be modeled as a Gaussian-distributed random variable with zero mean and a variance that follows \cite{macovski1983,wang2008,ma2012b}:

\begin{equation} \label{eq:noisemodel1}
\sigma_i^2=\frac{\exp(\bar{y_i})}{N_0^i} 
\end{equation}

\noindent In this equation, $\bar{y_i}$ is the expected value of the sinogram datum at detector $i$. In general, a system-specific constant $\eta$ is needed to fit the measurements \cite{wang2008}:

\begin{equation} \label{eq:noisemodel2}
\sigma_i^2= f_i \exp\left(\frac{\bar{y_i}}{\eta}\right)
\end{equation}

\noindent where $f_i$, similar to $1/N_0^i$ in \eqref{eq:noisemodel1}, mainly accounts for the effect of bowtie filtration.

\section{Image reconstruction algorithms in CT}

A central component in every CT system is the suite of image reconstruction and processing algorithms, whose task it to reconstruct the image of the object from its projection measurements. These algorithms have also continually evolved over time. The first CT scanners relied on simple iterative algorithms that aimed at recovering the unknown image as a solution of a system of linear equations. Many of these basic iterative methods had been developed by mathematicians like Kaczmarz well before the advent of CT. As the size of CT images grew, analytical filtered-backprojection (FBP) methods became more common and they are still widely used in practice \cite{pan2009}. For CBCT, the well-known Feldkamp-Davis-Kress (FDK) filtered-backprojetion algorithm is still widely used \cite{feldkamp1984, taguchi2004, tang2006}. These methods, which are based on the Fourier slice theorem, require a large number of projections to produce a high-quality image, but they are much faster than iterative methods.

The speed advantage of FBP methods has become less significant in recent years as the power of personal computers has increased and new hardware options such as graphical processing units (GPUs) have become available. On the other hand, with a consistent growth in medical CT usage, many studies have shown that the radiation dose levels used in CT may be harmful to the patients \cite{gonzalez2009, smith2009}. Reducing the radiation dose can be accomplished by reducing the number of projection measurements and/or by reducing the radiation dose for each projection. However, the images reconstructed from such under-sampled or noisy measurements with FBP methods will have a poor diagnostic quality. As a result of these developments, there has been a renewal of interest in statistical and iterative image reconstruction methods because they have the potential to produce high-quality images from low-dose scans \cite{beister2012, kalender2011}. Furthermore, even though in the beginning most of the algorithms used in CT were image reconstruction algorithms, gradually image processing algorithms were used for denoising, restoration, or otherwise improving the projection measurements and the reconstructed images. Many of these algorithms are borrowed from the research on image processing for natural images. Even today, algorithms that have been developed for processing of natural images are often applied in CT with little or no modifications.

\section{Patch-based methods}
\label{chapter:literature_review}

 In patch-based image processing, the units of operation are small image patches, which in the case of 3D images are also referred to as blocks. In the great majority of applications square patches or cubic blocks are used, even though other patch shapes can also be employed. For simplicity of presentation, we will use the term ``patch" unless when talking explicitly about 3D images. The number of pixels/voxels in a patch in patch-based image processing methods is usually on the order of tens or a few hundreds. A typical patch size would be $8 \times 8$ pixels for 2D images or $8 \times 8 \times 8$ voxels for 3D images.

Broadly speaking, in patch-based methods the image is first divided into small patches. Then, each patch is processed either separately on its own or jointly with patches that are very similar to it. The final output image is then formed by assembling the processed patches. In patch-based denoising, for instance, one can divide the image into small overlapping patches, denoise each patch independently, and then build the final denoised image using an averaging of the denoised patches. There are many reasons for focusing on small patches rather than on the whole image. First, because of the curse of dimensionality, it is much easier and more reliable to learn a model for small image patches than for very large patches or for the whole image. Secondly, for many models, computations are significantly reduced if they are applied on small patches rather than on the whole image. In addition, research in the past decade has shown that working with small patches can result in very effective algorithms that outperform competing methods in a wide range of image processing tasks. For example, as we will explain later in this chapter, patch-based denoising methods are currently considered to be the state of the art, achieving close-to-optimal denoising performance.

Patch-based methods have been among the most heavily researched methods in the field of image processing in recent years and they have produced state-of-the-art results in many tasks including denoising, restoration, super-resolution, inpainting, and reconstruction. However, these methods have received very little attention in CT. Even though there has been limited effort in using patch-based methods in CT, the results of the published works have been very promising. Given the great success of patch-based methods in various image-processing applications, they seem to have the potential to substantially improve the current state of the art algorithms in CT.

The word ``patch-based" may be ambiguous because it can potentially refer to any image model or algorithm that works with small patches. For instance, image compression algorithms such as JPEG work on small image patches. However, the word patch-based has recently been used to refer to certain classes of methods. In order to explain the central concepts of these methods, we will first describe the two main frameworks in patch-based image processing: (1) sparse representation of image patches in learned overcomplete dictionaries, (2) models based on nonlocal patch similarities. These two frameworks do not cover all patch-based image processing methods. However, most of these methods have their roots in one or both of these two frameworks.

\section{Image processing with learned overcomplete dictionaries}
\label{sec:DL}

\subsection{Sparse representation in analytical dictionaries}

A signal $x \in \mathbb{R}^m$ is said to have a sparse representation in a dictionary $D \in \mathbb{R}^{m \times n}$ if it can be accurately approximated by a linear combination of a small number of its columns. Mathematically, this means that there exists a vector $\gamma$ such that $x \cong D \gamma$ and $\|\gamma\|_0 \ll n$. Here, $\|\gamma\|_0$ denotes the the number of nonzero entries of $\gamma$ and is usually referred to as the $\ell_0$-norm of $\gamma$, although it is not a true norm. This means that only a small number of columns of $D$ are sufficient for accurate representation of the signal $x$. The ability to represent a high-dimensional signal as a linear combination of a small number of building blocks is a very powerful concept and it is at the center of many of the most widely used algorithms in signal and image processing. Columns of the dictionary $D$ are commonly referred to as atoms. If these atoms comprise a set of linearly independent vectors and if they span the whole space of $\mathbb{R}^m$, then they are called basis vectors and $D$ is called a basis. Moreover, if the basis vectors are mutually orthogonal, $D$ is called an orthogonal basis.

Bases, and orthogonal bases in particular, have interesting analytical properties that makes them easy to analyze. Moreover, for many of the orthogonal bases that are commonly used in signal and image processing, very fast computational algorithms have been developed. This computational advantage made these bases very appealing when the computational resources were limited. Over the past two decades, and especially in the past decade, there has been a significant shift of interest towards dictionaries that are adapted to a given class of signals using a learning strategy. The dictionaries obtained in this way lack the analytical and computational advantages of orthogonal bases, but they have much higher representational power. Therefore, they usually lead to superior results for many image processing tasks. Before we explain these dictionaries, we briefly review the history of sparsity-inducing transforms in image processing. More detailed treatment of this background can be found in \cite{mallat1999, rubinstein2010}.

Sparsity-based models are as old as digital signal processing itself. Starting in the 1960s, the Fourier transform was used in signal processing because it could diagonalize the linear time-invariant filters, which were widespread in signal processing. Adoption of the Fourier transform was significantly accelerated by the invention of the Fast Fourier Transform in 1965 \cite{cooley1965}. Fourier transform represents a signal as a sum of sinusoids of different frequencies. Suppressing the high-frequency components of this representation, for instance, is a simple denoising method. This is, however, not a good model for natural images because Fourier basis functions are not efficient for representing sharp edges. In fact, a single edge results in a large number of non-zero Fourier coefficients. Therefore, denoising using Fourier filtering leads to blurred images. An efficient representation of localized features needed bases that included elements with concentrated support. This gave rise to the Short-Time Fourier Transform (STFT) \cite{allen1977, bastiaans1980} and, more importantly, the wavelet transform \cite{daubechies1988, mallat1989}. The wavelet transform was the major manifestation of a revolution in signal processing that is referred to as multi-scale or multi-resolution signal processing. The main idea in this paradigm is that many signals, and in particular natural images, contain relevant features on many different scales. Both the Fourier transform and the wavelet transform can be interpreted as the representation of a signal in a dictionary. For the Fourier transform, for example, the dictionary atoms include sinusoids of different frequencies.

Despite its tremendous success, the wavelet transform suffers from important shortcomings for analyzing higher-dimensional signals such as natural images. Even though wavelet transform possesses important optimality properties for one-dimensional signals, it is much less effective for higher-dimensional signals. This is because in higher dimensions, wavelet transform is a separable extension of the one-dimensional transform along different dimensions. As a result, for example the 2D wavelet transform is suitable for representing points but it is not effective for representing edges. This is a major shortcoming because the main features in natural images are composed of edges. Therefore, there was a need for sparsity-inducing transforms or dictionaries that could efficiently represent these types of features. Consequenbtly, great research effort was devoted to designing transforms/dictionaries especially suitable for natural images. Among the proposed transforms, some of them that have been more successful for image processing applications include the complex wavelet transform \cite{kingsbury2001}, the curvelet transform \cite{candes2000, candes2006b}, the contourlet transform \cite{do2005} and its extension to 3D images known as surfacelet \cite{lu2007}, the shearlt transform \cite{labate2005, easley2008}, and the bandlet transform \cite{lepennec2005}.

The transforms mentioned above have had a great impact on the field of image processing and they are still used in practice. They have also been used in CT \citep [e.g.,][]{rantala2006, candes2000, frikel2013, vandeghinste2013}. However, learned overcomplete dictionaries achieve much better results in practice by breaking some of the restrictions that are naturally imposed by these analytical dictionaries. The restriction of orthogonality, for instance, requires the number of atoms in the dictionary to be no more than the dimensionality of the signal. The consequences of these limitations had already been realized by researchers working on wavelets. This realization led to developments such as stationary wavelet transform, steerable wavelet transform, and wavelet packets, which greatly improved upon the orthogonal wavelet transform \cite{simoncelli1992, nason1995, coifman1992}. However, these transforms are still based on fixed constructions and do not have the freedom and adaptability of learned dictionaries that we will explain below.

\subsection{Learned overcomplete dictionaries}
\label{sec:DL_learning}

The basic idea of adapting the dictionary to the signal is not completely new. One can argue that the Principal Component Analysis (PCA) method \cite{hotteling1993}, which is also known as the Karhunen--Loève Transform (KLT) in signal processing, is an example of learning a dictionary from the training data. However, this transform too is limited in terms of the dictionary structure and the number of atoms in the dictionary. Specifically, the atoms in a PCA dictionary are necessarily orthogonal and their number is at most equal to the signal dimensionality.

The modern story of dictionary learning begins with a paper by Olshausen and Field \cite{olshausen1996}. The question posed in that paper was: if we assume that small patches of natural images have a sparse representation in a dictionary $D$ and try to learn this dictionary from a set of training patches, what would the learned dictionary atoms look like? They found that the learned dictionary consisted of atoms that were spatially localized, oriented, and bandpass. This was a remarkable discovery because these are exactly the characteristics of simple-cell receptive fields in the mammalian visual cortex. Although similar patterns existed in Gabor filters \cite{daugman1980, daugman1985}, Olshausen and Field had been able to show that these structures can be explained using only one assumption: sparsity.   

Suppose that we are given a set of training signals and would like to learn a dictionary for sparse representation of these signals. We stack these training signals as columns of a matrix, which we denote with $X$. Each column of $X$ is a referred to as a training signal. In image processing applications, each training signal is a patch (for 2D images) or block (in the case of 3D images) that is vectorized to form a column of $X$. Using the matrix of training signals, a dictionary can be learned through the following optimization problem.

\begin{equation} \label{eq:DL_olshausen}
\operatorname*{\textbf{minimize}}_{D \in \mathcal{D},\Gamma} \|X - D \Gamma \|_F^2 + \lambda \|\Gamma \|_1
\end{equation}

In the above equation, $X$ denotes the matrix of training data, $\Gamma$ is the matrix of representation coefficients of the training signals in $D$, and $\mathcal{D}$ is the set of matrices whose columns have a unit Euclidean norm. The $i^{\text{th}}$ column of $\Gamma$ is the vector of representation coefficients of the $i^{\text{th}}$ column of $X$ (i.e.,  the $i^{\text{th}}$ training signal) in $D$. The notations $\|.\|_F$ and $\|.\|_1$ denote, respectively, the Frobenius norm and the $\ell_1$ norm. The constraint $D \in \mathcal{D}$ is necessary to avoid scale ambiguity because without this constraint the objective function can be made smaller by decreasing $\Gamma$ by an arbitrary factor and increasing $D$ by the same factor. The first term in the objective function requires that the training signals be accurately represented by the columns of $D$ and the second terms promotes sparsity, encouraging that a small number of columns of $D$ are used in the representation of each training signal.

There are many possible variations of the optimization problem presented in Equation \eqref{eq:DL_olshausen}, some of which we will explain in this chapter. For example the $\ell_1$ penalty on $\Gamma$ is sometimes replaced with an $\ell_0$ penalty. In fact, it can be shown that variations of this problem include problems as diverse as PCA, clustering or vector quantization, independent component analysis, archetypal analysis, and non-negative matrix factorization (see for example \cite{bach2012, mairal2014}). The most important fact about the optimization problem in \eqref{eq:DL_olshausen} is that it is not jointly convex with respect to $D$ and $\Gamma$. Therefore, only a stationary point can be hoped for and the global optimum is not guaranteed. However, this problem is convex with respect to $D$ and $\Gamma$ individually. Therefore, many dictionary learning problems adopt an alternating minimization approach. In other words, the objective function is minimized with respect to one of the two variables while keeping the other fixed. The first such method was the method of optimal directions (MOD) \cite{engan1999}. In each iteration of MOD, the objective function is first minimized with respect to $\Gamma$ by solving a separate sparse coding problem for each training signal:

\begin{equation} \label{eq:sparse_coding_MOD}
\Gamma_i^{k+1}= \operatorname*{\textbf{argmin}}_{\gamma} \|X_i - D^k \gamma \|_2^2 \hspace{0.5cm} \textbf{subject to:} \hspace{0.5cm} \|\gamma \|_0 \leq K
\end{equation}

\noindent In the above equation, and in the rest of this chapter, we use subscripts on matrices to index their columns. Therefore, $X_i$ indicates the $i^{\text{th}}$ column of $X$, which is the $i^{\text{th}}$ training signal and $\Gamma_i$ is the $i^{\text{th}}$ column of $\Gamma$, which is the vector of representation coefficients of $X_i$ in $D$. We will use superscripts to indicate iteration number. Once all columns of $\Gamma$ are updated, $\Gamma$ is kept fixed and the dictionary is updated. This update is in the form of a least-squares problem that has a closed-form solution:

\begin{equation} \label{eq:dictionary_update_MOD}
D^{k+1}= X (\Gamma_i^{k+1})^\dagger
\end{equation}

\noindent where $\dagger$ denotes the Moore-Penrose pseudo-inverse.

Before moving on, we need to say two brief words about the optimization problem in \eqref{eq:sparse_coding_MOD}. This optimization problem is one formulation of the sparse coding problem that is a central part of any image processing method that makes use of learned overcomplete dictionaries. Because of their ubiquity, there has been a very large body of research on the properties of these problems and solution methods. We will only mention or describe the relevant algorithms where necessary. A recent review of these methods can be found in \cite{bach2012}. In MOD, this step is solved using the orthogonal matching pursuit (OMP) \cite{pati1993} or the focal underdetermined system solver (FOCUSS) \cite{gorodnitsky1997}.

Another dictionary-learning algorithm that has shown to be more efficient than MOD is the K-SVD algorithm \cite{aharon2006}. K-SVD is arguably the most widely used dictionary learning algorithm today. Similar to MOD, each iteration of the K-SVD algorithm updates each column of $\Gamma$ by solving a sparse coding problem similar to \eqref{eq:sparse_coding_MOD}. However, unlike the MOD that updates all dictionary atoms at once, K-SVD updates each dictionary atom (i.e., each column of $D$) sequentially. Assuming all dictionary atoms are fixed except for the $i^{\text{th}}$ atom, the cost function in \eqref{eq:DL_olshausen} can be written as:

\begin{equation} \label{eq:dictionary_update_KSVD}
\begin{aligned}
\|X-D \Gamma \|_F^2 &= \left\| X- \sum_{j=1}^N D_j \Gamma_j^T \right\|_F^2 \\
&= \left\|X- \sum_{j=1, j \neq i }^N D_j \Gamma_j^T - D_i \Gamma_i^T \right\|_F^2 = \|E_i - D_i \Gamma_i^T\|_F^2
\end{aligned}
\end{equation}

\noindent In the K-SVD algorithm this is minimized using an SVD decomposition of the matrix $E_i$ after restricting it to the training signals that are using $D_i$ in their representation. The reason behind this restriction is that it will preserve the sparsity of the representation coefficients. Let us denote the restricted version of $E_i$ with $E_i^R$ and assume that the SVD decomposition of $E_i^R$ is $E_i^R= U \Delta V^T$. Then, $U_1$ and $\Delta(1,1) V_1$ provide the updates of $D_i$ and $\Gamma_i^T$, where $U_1$ and $V_1$ denote the first columns of $U$ and $V$, respectively, and $\Delta(1,1)$ is the largest singular value of $E_i^R$.

A major problem with methods like MOD and K-SVD is that they are computationally intensive. Even though efficient implementations of these algorithms have been developed \cite{rubinstein2008}, computations become very excessive when the number of training signals and the signal dimensionality grow. Therefore, a number of studies have proposed algorithms that are particularly designed for learning dictionaries from huge datasets in reasonable time \cite{mairal2010, thom2015}. The algorithm proposed in \cite{mairal2009b, mairal2010}, for instance, is based on stochastic optimization algorithms that are particularly suitable for large-scale problems. Instead of solving the optimization problem by considering the whole training data, it randomly picks one training signal (i.e., one column of $X$) and approximately minimizes the objective function using that one training signal. Convincing theoretical and empirical evidence regarding the convergence of this dictionary learning approach have been presented in \cite{mairal2010}. 

Another important class of dictionary learning algorithms are maximum- likelihood algorithms, which are in fact among the first methods suggested for learning dictionaries from data \cite{lewicki1999, lewicki2000, kreutz2003}. These methods assume that each training signal is produced by a model of the form:

\begin{equation} \label{eq:signal_model_probabilistic}
X_i= D \Gamma_i + w_i
\end{equation}

\noindent where $w_i$ is a Gaussian-distributed white noise. To encourage sparsity of the representation coefficients ($\Gamma_i$), these methods assume a sparsity-promoting prior such as a Cauchy or Laplace distribution for entries of $\Gamma$. Additionally, these approaches assume that the entries of $\Gamma$ are independent and identically distributed and that each signal $X_i$ is drawn independently. A dictionary can then be learned by maximizing the data likelihood $p(X|D)$ or the posterior $p(D|X)$. Quite often, the resulting likelihood function is very difficult to maximize and it is further simplified before applying the optimization algorithm.

It can be argued that the maximum-likelihood methods explained above are not truly Bayesian methods because they yield a point estimate rather than the full posterior distribution \cite{mairal2014}. As a result, in recent years several fully-Bayesian dictionary learning methods have been proposed \cite{zhou2009, zhou2012, hughes2013}. In these algorithms, priors are placed on all model parameters, i.e., not only the dictionary atoms $D_i$ and sparse representation vectors $\Gamma_i$, but also on all other model parameters such as the number of dictionary atoms and the noise variance for each training signal. The most important priors assumed in these models are usually Gaussian priors with Gamma hyper-priors for the dictionary atoms ($D_i$) and representation coefficients ($\Gamma_i$), and a Beta-Bernoulli process for the support of $\Gamma_i$ \cite{zhou2009, hughes2013}. Full posterior density of the model parameters and hyper-parameters are iteratively estimated via Gibbs sampling. Compared to all other dictionary learning methods described above, these fully-Bayesian methods are significantly more computationally demanding. On the other hand, their robustness with respect to poor initialization and their ability to learn some important parameters such as the noise variance makes them potentially very useful for certain applications \cite{xing2012, stevens2014}.

There are many variations and enhancements of dictionary learning that we cannot describe in detail due to space limitations. However, we briefly mention three important variations. The first is the structured dictionary learning. The main idea here is not only to learn the dictionary atoms but also the interaction between the learned dictionary atoms. For example, a common structure that is assumed between the atoms is a tree structure, where each atom is the descendant/parent of some other atoms \cite{jenatton2010, jenatton2011}. During the dictionary usage, then, an atom will participate in the sparse code of a signal if and only if its parent atom does so. Obviously, the basic $\ell_0$ and $\ell_1$ norms are not capable of modelling these interactions between dictionary atoms. The success of structured dictionary learning, therefore, has been made possible by algorithms for structured sparse coding \cite{huang2011b, jenatton2011b}. Another common structure is the grid structure that enforces a neighborhood relation between atoms \cite{kavukcuoglu2009, thom2015}. The second variation that is of great importance is multi-scale dictionary learning. Extending the basic dictionary learning scheme to consider different patch sizes has been shown to significantly improve the performance of the dictionary-based image processing \cite{mairal2008c, ophir2011}. Moreover, this extension to multiple scales has been suggested as an approach to addressing some of the theoretical flaws in the dictionary-based image processing \cite{papyan2016}. The third important variation that we mention here includes dictionaries that have a fast application. As we mentioned above, learned dictionaries do not possess such desired structural properties as orthogonality. As a result, they are much more costly to apply than analytical dictionaries. Therefore, several dictionary structures have been proposed with the goal of reducing the computational cost during dictionary usage \cite{rubinstein2010b, aharon2008}. These dictionaries can be particularly useful for processing of large 3D images.

As final remarks on dictionary learning, we should first mention that there is no strong theoretical justification behind most dictionary learning algorithms. In particular, there is no theoretical guarantee that these algorithms are robust or that the learned dictionary should work well in practical applications. In practice, learning of a good dictionary certainly requires sufficient amount of training data and the minimum amount of data needed grows at least linearly with the number of dictionary atoms \cite{rubinstein2009, rubinstein2010b}. Uniqueness of the learned dictionary, however, is only guaranteed for an exponential number of training signals \cite{aharon2006b}. In fact, the theory of dictionary learning is considered to be one of the major open problems in the field of sparse representation \cite{elad2012}. Secondly, pre-processing of training image patches has proved to significantly influence the types of structures that emerge in the learned dictionary and the performance of the learned dictionary in practice. Three of the most commonly used pre-processing operations include: (i) removing of the patch mean, also known as centering \cite{elad2006}, (ii) variance normalization which is preceded with centering \cite{jarrett2009, pinto2008}, and (iii) de-correlating the pixel values within a patch, referred to as whitening \cite{bell1997, hyvarinen2009}. The overall effect of all these three operations is to amplify the high-frequency structure such as edges, resulting in more high-frequency patterns in the learned dictionary \cite{mairal2014}.

\subsection{Applications of learned dictionaries}
\label{sec:DL_applications}

Learned overcomplete dictionaries have been employed in various image processing and computer vision applications in the past ten years. There are monographs that review and explain these applications in detail \cite{mairal2014, elad2010}. Because of space limitations, we describe the basic formulations for image denoising, image inpainting, and image scale-up. Not only these three tasks are among the most successful applications of learned dictionaries in image processing, they are also very instructive in terms of how these dictionaries can be used to accomplish various image processing tasks.

\begin{description}

\item[Image denoising] \

Suppose that we have measured a noisy image $x= x_0+w$, where $x_0$ is the true underlying image and $w$ is the additive noise that is assumed to be white Gaussian. The prior assumption in denoising using a dictionary $D$ is that every patch in the image has a sparse representation in $D$. If we denote a typical patch with $p$, this would mean that there exists a sparse vector $\gamma$ such that $||p-D \gamma||_2^2 < \epsilon$, where $\epsilon$ is proportional to the noise variance \cite{elad2006, mairal2008}. Using this prior on every patch in the image, the maximum a posteriori (MAP) estimation of the true image can be found as the solution of the following problem \cite{elad2010}:

\begin{equation} \label{eq:KSVD_denoising_problem}
\Big\{ \widehat{x_0}, \{\widehat{\gamma}_i \}_{i=1}^N \Big\}= \operatorname*{\textbf{argmin}}_{z, \{\gamma_i \}_{i=1}^N} \lambda \|z-x \|_2^2 + \sum_{i=1}^N \big( \|R_iz- D\gamma_i\|_2^2+ \|\gamma_i\|_0 \big)
\end{equation}

\noindent where $R_i$ represents a binary matrix that extracts and vectorizes the $i^{\text{th}}$ patch from the image. This is a very common notation. $N$ is the total number of extracted patches. It is common to use overlapping patches to avoid discontinuity artifacts at the patch boundaries. In fact, unless the computational time is a concern, it is recommended that maximum overlap is used such that adjacent extracted patches are shifted by only one pixel in each direction. This means extracting all possible patches from the image.

The objective function in Equation \eqref{eq:KSVD_denoising_problem} is easy to understand. The first term requires the denoised image to be close to the measurement, $x$, and the second term requires that every patch extracted from this image to have a sparse representation in the dictionary $D$. The common approach to solving this optimization problem is an approximate block-coordinate minimization. First, we initialize $z$ to the noisy measurement ($z= x$). Keeping $z$ fixed, the objective function is minimized with respect to $\{\gamma_i \}_{i=1}^N$. This step is simplified because it is equivalent to $N$ independent problems, one for each patch, that can be solved using sparse coding algorithms. Then $\{\gamma_i \}_{i=1}^N$ are kept fixed and the objective function is minimized with respect to $z$. This minimization has a closed-form solution:

\begin{equation} \label{eq:KSVD_denoising_update}
\widehat{x_0}=  \Bigg( \lambda I + \sum_{i=1}^N R_i^T R_i \Bigg)^{-1} \Bigg( \lambda x + \sum_{i=1}^N R_i^T D \hat{\gamma_i} \Bigg)
\end{equation}

There is no need to form and invert a matrix to solve this equation. It is basically equivalent to returning the denoised patches to their right place on the image canvas and performing a weighted averaging. The weighted averaging simply takes into account the overlapping of the patches and a weighted averaging with the noisy image $x$ (with weight $\lambda$).

The minimization with respect to $\{\gamma_i \}_{i=1}^N$ and $z$ can be performed iteratively by using $\widehat{x_0}$ obtained from \eqref{eq:KSVD_denoising_update} as the new estimate of the image. However, this will run into difficulties because the noise distribution in $\widehat{x_0}$ is unknown and it is certainly not white Gaussian. Therefore, $\widehat{x_0}$ obtained from \eqref{eq:KSVD_denoising_update} is usually used as the estimate of the underlying image $x_0$.

\item[Image inpainting] \

Let us denote the true underlying image with $x_0$ and assume that the observed image $x$ not only contains noise, but also some pixels are not observed or are corrupted to the extent that the measurements of those pixels should be ignored. The model used for this scenario is $x= M x_0 + w$ where $w$ is the additive noise and $M$ is a mask matrix, which is a binary matrix that removes the unobserved/corrupted pixels. The goal is to recover $x_0$ from x. Similar to the denoising problem above, we use the prior assumptions that patches of $x_0$ have a sparse representation in a dictionary $D$. The MAP estimate of $x_0$ can be found as a solution of the following problem \cite{elad2010}:

\begin{equation} \label{eq:inpainting_problem}
\begin{aligned}
\Big\{ \widehat{x_0}, \{\widehat{\gamma}_i \}_{i=1}^N \Big\}= \operatorname*{\textbf{argmin}}_{z, \{\gamma_i \}_{i=1}^N} &\lambda \|Mz-x \|_2^2  \\ &+  \sum_{i=1}^N \big( \|R_iz- D\gamma_i\|_2^2+ \|\gamma_i\|_0 \big)
\end{aligned}
\end{equation}

An approximate solution can be found using an approach rather similar to that described above for the denoising problem. Specifically, we start with an initialization $z= M^Tx$. Then, assuming that $z$ is fixed, we solve $N$ independent sparse coding problems to find estimates of $\{\gamma_i \}_{i=1}^N$. The only issue here is that this initial $z$ will be corrupted at the locations of unobserved pixels. Therefore, the estimation of $\{\gamma_i \}_{i=1}^N$ needs to take this into account by introducing a local mask matrix for each patch:

\begin{equation} \label{eq:sparse_coding_inpainting}
\hat{\gamma_i}= \operatorname*{\textbf{argmin}}_{\gamma} \|M_i(R_i z - D \gamma) \|_2^2 \hspace{0.5cm} \textbf{subject to:} \hspace{0.5cm} \|\gamma \|_0 \leq K
\end{equation}

Once $\{\gamma_i \}_{i=1}^N$ are estimated, an approximation to the underlying full image is found as:

\begin{equation} \label{eq:inpainting_update}
\hat{x_0}=  \Bigg( \lambda M^TM + \sum_{i=1}^N R_i^T R_i \Bigg)^{-1} \Bigg( \lambda M^Tx + \sum_{i=1}^N R_i^T D \hat{\gamma_i} \Bigg)
\end{equation}

\noindent which has a simple interpretation similar to \eqref{eq:KSVD_denoising_update}. The method described above has been shown to be very effective in many studies \cite{peyre2009, mairal2008, mairal2008c}.

\item[Image scale-up (super-resolution)] \

As we saw above, the applications of learned dictionaries for image denoising and inpainting can be quite straightforward. Nevertheless, application of learned dictionaries for image processing may involve much more elaborate approaches, even for the simple tasks such as denoising. As an example of a slightly more complex task, in this section we explain the image scale-up. Image scale-up can serve as a good example of more elaborate applications of learned dictionaries in image processing. Moreover, it has been one of the most successful applications of learned dictionaries to date \cite{yang2010}.

Suppose $x_h$ is a high-resolution image. A blurred low-resolution version of this image can be modeled as $x_l= SHx_h$, where $H$ and $S$ denote the blur and down-sampling operators. Given the measured low-resolution image, which can also include additive noise (i.e., $x_l= SHx_h+ w$), the goal is to recover the high-resolution image. This problem is usually called the image scale-up problem, and it is also referred to as image super-resolution.

The first image scale-up algorithm that used learned dictionaries was suggested in \cite{yang2010}. This algorithm is based on learning two dictionaries, one for sparse representation of the patches of the high-resolution image and one for sparse representation of the patches of the low-resolution image. Let us denote these dictionaries with $D_h$ and $D_l$, respectively. The basic assumption in this algorithm is that sparse representation of a low-resolution patch in $D_l$ is identical to the sparse representation of its corresponding high-resolution patch in $D_h$. Therefore, given a low-resolution image $x_l$, one can divide it into patches and use each low-resolution patch to estimate its corresponding high-resolution patch. Let us denote the $i^{\text{th}}$ patch extracted from $x_l$ with $X^l_i$ and its corresponding high-resolution patch with $X^h_i$. One first finds the sparse representation of $X^l_i$ in $D_l$ using any sparse coding algorithm such that $X^l_i\cong D \gamma_i$. Then, by assumption, $\gamma_i$ is also the sparse representation of $X^h_i$ in $D_h$. Therefore, the estimate of $X^h_i$ will be: $\widehat{X^h_i}\cong D_h \gamma_i$. These estimated high-resolution patches are then placed on the canvas of the high-resolution image and the high-resolution image is formed via a weighted averaging similar to that in the denoising application above. The procedure that we explained here for estimating the high-resolution patches from their low-resolution counterparts is the simplest approach. In practice, this procedure is applied with slight modifications that significantly improve the results \cite{yang2010, zeyde2012, elad2010}.

The main assumption in the above algorithm was that the sparse codes of the low-resolution and high-resolution patches were identical. This is an assumption that has to be enforced during dictionary learning. In other words, the dictionaries $D_h$ and $D_l$ are learned such that this condition is satisfied. The dictionary learning approach suggested in \cite{yang2010} is:

\begin{equation} \label{eq:DL_upscaling}
\operatorname*{\textbf{minimize}}_{D_h, D_l, \Gamma} \frac{1}{m_h} \|X^h - D_h \Gamma \|_F^2 + \frac{1}{m_l} \|X^l - D_l \Gamma \|_F^2 + \lambda \|\Gamma \|_1
\end{equation}

\noindent where $X^h$ and $X^l$ represent the matrices of training signals. The  $i^{\text{th}}$ column of $X^h$ is the vectorized version of a patch extracted from a high-resolution image and the $i^{\text{th}}$ column of $X^l$ is the vectorized version of the corresponding low-resolution patch.  $m_h$ and $m_l$ are the lengths of the high-resolution and low-resolution training signals and are included in the objective function to properly balance the two terms. The important choice in the objective function in \eqref{eq:DL_upscaling} is to use the same $\Gamma$ in the first and the second terms of the objective function. It is easy to understand how this choice forces the learned dictionaries $D_h$ and $D_l$ to be such that the corresponding high-resolution and low-resolution patches have the same sparse representation.

The above algorithm achieved surprisingly good results \cite{yang2010}. However, it was soon realized that the assumption of this algorithm on the sparse representations was too restrictive and that better results could be obtained by relaxing those assumptions. For instance, one study suggested a linear relation between the sparse representations of low-resolution and high-resolution patches and obtained better results \cite{wang2012}. The dictionary learning formulation for this algorithm had the following form:

\begin{equation} \label{eq:DL_upscaling_semicoupled}
\begin{aligned}
& \operatorname*{\textbf{minimize}\,}_{\left\lbrace D_h,\Gamma^h,D_l,\Gamma^l, W \right\rbrace} \quad \bigg( ||X^h- D_h \Gamma^h||_F^2 + ||X^l- D_l \Gamma^l||_F^2 \\
& \hspace{95pt} + \lambda_h ||\Gamma^h||_1 + \lambda_l ||\Gamma^l||_1 \\
& \hspace{95pt} + \lambda_W ||\Gamma^l - W \Gamma^h||_F^2 + \alpha ||W||_F^2  \bigg) 
\end{aligned}
\end{equation}

It is easy to see that here the assumption is not that the sparse representation of high-resolution patches ($\Gamma^h$) is the same as the sparse representation of the low-resolution patches ($\Gamma^l$), but that there is a linear relationship between them. This linear relation is represented by the matrix $W$. This results in a much more general and more flexible model. On the other hand, this is also a more difficult model to learn because it requires learning of the matrix $W$, in addition to the two dictionaries. In \cite{wang2012}, a block-coordinate optimization algorithm was suggested for solving this problem and it was shown to produce very good results.

There have also been other approaches to relaxing the relationship between the sparse codes of high-resolution and low-resolution patches. For instance, one study suggested a bilinear relation involving two matrices \cite{huang2013}. Another study suggested a statistical inference technique to predict the sparse code of the high-resolution patches from low-resolution ones \cite{peleg2014}. Both of these approaches reported very good results. In general, image scale-up with the help of learned dictionaries has shown to outperform other competing methods and it is a good example of the power of learned dictionaries in modeling natural images.

\item[Other applications] \

In the above, we explained three applications of learned dictionaries. However, these dictionaries have proved highly effective in many other applications as well. Some of these other applications include image demosaicking \cite{mairal2008, menon2011}, deblurring \cite{couzinie2011, dong2011}, compressed sensing \cite{rauhut2008, duarte2009, chen2013b}, morphological component analysis \cite{elad2010}, compression \cite{bryt2008, skretting2011}, classification \cite{jiang2013, ramirez2010}, cross-domain image synthesis \cite{zhuang2013b} and removal of various types of artifacts from the image \cite{yeh2014, li2012d}.

For many image processing tasks, such as denoising and compression, application of the dictionary is relatively straightforward. However, there are also more complex tasks for which learning and application of overcomplete dictionaries is much more complex. It has been suggested in \cite{mairal2012, mairal2014} that many of these applications can be considered as instances of classification or regression problems. The authors of \cite{mairal2012} coin the term ``task-driven dictionary learning" to describe these applications and suggest that a general optimization formulation for these problems if of this form:

\begin{equation} \label{eq:general_task_driven}
\operatorname*{\textbf{minimize}\,}_{D \in \mathcal{D}, W \in \mathcal{W}}  \mathcal{L}(Y,W,\hat{\Gamma})+ \lambda \|W \|_F^2
\end{equation}

In the above equation, $\hat{\Gamma}$ is the matrix of representation coefficients of the training signals, obtained by solving a problem such as \eqref{eq:sparse_coding_MOD}. The cost function $\mathcal{L}$ quantifies the error in the prediction of the target variables $Y$ from the sparse codes $\hat{\Gamma}$, and $W$ denotes the model parameters. For a classification problem $Y$ represents the labels of the training signals, whereas in a regression setting $Y$ represents real-valued vectors. For instance, the image scale-up problem that we presented above is an example of the regression setting where $Y$ represents the vectors of pixel values of the high-resolution patches. The second term in the above objective function is  a regularization term on model parameters that is meant to avoid overfitting and numerical instability.

Therefore, in task-driven dictionary learning the goal is to learn the dictionary not only for sparse representation of the signal, but also so that it can be employed for accurate prediction of the target variables, $Y$. The general optimization problem in \eqref{eq:general_task_driven} is very difficult to solve. In addition to the fact that the objective function is non-convex, the dependence of $\mathcal{L}$ on $D$ is through $\hat{\Gamma}$, which is in turn obtained by solving \eqref{eq:sparse_coding_MOD}. In the asymptotic case when the amount of training data is very large, it has been shown that this general optimization problem is differentiable and can be effectively solved using stochastic gradient descent \cite{mairal2012}. It has been shown that this approach can lead to very good results in a range of classification and regression tasks such as compressed sensing, handwritten digit classification, and inverse
halftoning \cite{mairal2012}.

\end{description}

\section{Non-local patch-based image processing}
\label{sec:NLM}

Natural images contain abundant self-similarities. Speaking in terms of patches, this means that for every patch in a natural image we can probably find many similar patches in the same image. The main idea in non-local patch-based image processing is to exploit this self-similarity by finding/collecting similar patches and processing them jointly. The idea of exploiting patch similarities and the notion of nonlocal filtering are not very new \cite{efros1999, criminisi2004, wei2000, tomasi1998, yaroslavsky1985}. However, it was the non-local means (NLM) denoising algorithm proposed in \cite{buades2005} that started the new wave of research in this field. Even though the basic idea behind NLM denoising is very simple and intuitive, it achieves remarkable denoising results and it has created a great deal of interest in the image processing community.

Let us denote the noisy image with $x= x_0+w$, where, as before, $x_0$ denotes the true image. We also denote the $i^{\text{th}}$ pixel of $x$ with $x(i)$ and a patch/block centered on $x(i)$ with $x[i]$. The NLM algorithm considers overlapping patches, each patch centered on one pixel. The value of the $i^{\text{th}}$ pixel in the underlying image, $x_0(i)$, is estimated as a weighted average of the center pixels of all the patches as follows:

\begin{equation} \label{eq:nlm}
\widehat{x_0}(i)= \sum_{j=1}^N \frac{G_a(x[j]-x[i])}{\sum_{j=1}^N G_a(x[j]-x[i])} x(j)
\end{equation}

\noindent where $G_a$ denotes a Gaussian kernel with bandwidth $a$ and $N$ is the total number of patches. The intuition behind this algorithm is very simple: similar patches are likely to have similar pixels at their centers. Therefore, in order to estimate the true value of the $i^{\text{th}}$ pixel, the algorithm performs a weighted averaging of the values of all pixels, with the weight being related to the similarity of each patch with the patch centered on the $i^{\text{th}}$ pixel. Although in theory all patches can be included in the denoising of the $i^{\text{th}}$ pixel, as shown in \eqref{eq:nlm}, in practice only patches from a small neighborhood around this pixel are included. In fact, many of the methods that are based on NLM denoising first find several patches that are similar to $x[i]$. Only those patches that are similar enough to $x[i]$ are used in computing $\widehat{x_0}(i)$. Therefore, a practical implementation of the NLM denoising will be:

\begin{equation} \label{eq:nlm_practical}
\begin{aligned}
\widehat{x_0}(i)= & \sum_{j \in S_i} \frac{G_a(x[j]-x[i])}{\sum_{j \in S_i} G_a(x[j]-x[i])} x(j)   \hspace{0.5cm} \\ &  \textbf{where:}   \hspace{0.5cm} S_i= \{j| \; j \in \mathcal{N}_i  \;  \&  \;  \| x[j]-x[i] \|_2 \leq \epsilon  \}
\end{aligned}
\end{equation}

\noindent where $\mathcal{N}_i$ is a small neighborhood around the $i^{\text{th}}$ pixel and $\epsilon$ is a noise-dependent threshold.

The idea behind the NLM has proved to be an extremely powerful model for natural images. For the denoising task, NLM filtering and its extensions have led to the best denoising results \cite{milanfar2013, shao2014}. Some studies have shown that the current state-of-the-art algorithms are approaching the theoretical performance limits of denoising \cite{levin2011, chatterjee2012}. Some of the recent extensions of the basic NLM denoising include Bayeian/probabilistic extensions of the method \cite{lebrun2013, wu2013}, spatially adaptive selection of the algorithm parameters \cite{dore2009, duval2011}, combining NLM denoising with TV denoising \cite{sutour2014}, and the use of non-square patches that has been shown to improve the results around edges and high-contrast features \cite{deledalle2012b}. Some of the most productive extensions of the NLM scheme involve exploiting the power of learned dictionaries. We will discuss these methods in the next section.

Nonlocal patch-based methods are very computationally demanding. Therefore, a large number of research papers have focused on speedup strategies. A very effective strategy was proposed in \cite{darbon2005b}. This strategy is based on building a temporary image that holds the discrete integration of the squared differences of the noisy image for all patch translations. This integral image is used for fast computation of the patch differences ($x[j]-x[i]$), which is the main computational burden in nonlocal patch-based methods. A large number of paper have focused on reducing the computational cost of NLM denoising by classifying/clustering the image patches before the start of the denoising \cite{mahmoudi2005, dauwe2008, bhujle2014}. The justification behind this approach is that the computational bottleneck of NLM denoising is the search for similar patches. Therefore, these methods aim at clustering the patches so that the search for similar patches becomes less computationally demanding. Most of these methods compute a few features from each patch to obtain a concise representation of the patches. Typical features include average gray value and gradient orientation. During denoising, for each patch a set of similar patches is found using the clustered patches. One study compared various tree structures for fast finding of similar patches in an image and found that vantage point trees are superior to other tree structures \cite{kumar2008}. Another class of highly efficient algorithms for finding similar patches are stochastic in nature. These methods can be much faster than the deterministic techniques we mentioned above, but they are less accurate. Perhaps the most widely used algorithm in this category is the PatchMatch algorithm and its extensions \cite{barnes2009, barnes2010}.

The NLM algorithm and its extensions that we will explain in the next section have been recognized as the state-of-the-art methods for image denoising. However, the idea of exploiting the patch similarities has been used for many other image processing tasks. For instance, it has been shown that nonlocal patch similarities can be used to develop highly effective regularizations for inverse problems and iterative image reconstruction algorithms \cite{mohsin2015, peyre2008, gilboa2006, lou2010, zhang2010b}. Below, we briefly explain two of these algorithms.

Let us consider the inverse problem of estimating an unknown image $x$ from the measurements $y= Ax+w$, where $w$ is the additive noise. The matrix $A$ represents the known forward model that can be, for example, a blur matrix (in image deblurring) or the projection matrix (in tomography). In \cite{peyre2008}, it is suggested to recover $x$ by solving this optimization problem:

\begin{equation} \label{eq:peyre_inverse_problem}
\hat{x}= \operatorname*{\textbf{argmin}}_{x} \|y- Ax \|_2^2 + \lambda \sum_i \sum_j \sqrt{w_{i,j}} |(x(i)-x(j)|
\end{equation}

\noindent where $w_{i,j}$ are the nonlocal patch-based weights that are computed similar to NLM denoising:

\begin{equation} \label{eq:peyre_weights}
w_{i,j}= \frac{1}{Z} \exp{ \bigg(-\frac{\|x[i]-x[j] \|}{2 \sigma^2} \bigg)}
\end{equation}

\noindent where $Z$ is a normalizing factor. 
Therefore, the regularization term in \eqref{eq:peyre_inverse_problem} is a non-local total variation on a graph where the graph weights are based on nonlocal patch similarities. The difficulty with solving this optimization problem is that the weights themselves depend on the unknown image, $x$. The algorithm suggested in \cite{peyre2008} iteratively estimates the weights from the latest image estimate and then updates the image based on the new weights using a proximal gradient method. In summary, given the image estimate at the $k^{\text{th}}$ iteration, $\hat{x}^k$, the weights are estimated from this image. Then, the image is updated using a proximal gradient iteration \cite{chambolle2004, combettes2011} :

\begin{equation} \label{eq:peyre_prox_gradient}
\begin{aligned}
\hat{x}^{k+1}= & \textbf{Prox}_{\mu \lambda J} \Bigg( \hat{x}^k+ \mu A^T \big( y- A \hat{x}^k \big)   \Bigg)  \\
& \textbf{Prox}_{J}(x) = \operatorname*{\textbf{argmin}}_{z} \frac{1}{2} \| x- z \|_2^2 + J(z)
\end{aligned}
\end{equation}

\noindent where $J$ is the regularization term in \eqref{eq:peyre_inverse_problem} and $\mu$ is the step size. Having computed the new estimate $\hat{x}^{k+1}$, the patch-based weights are re-computed and the algorithm continues. This algorithm showed very good results on three types of inverse problems including compressed sensing, inpainting, and image scale-up \cite{peyre2009}.

In \cite{yang2013}, the following optimization problem was suggested for recovering the unknown image $x$. 

\begin{equation} \label{eq:yang_inverse_problem}
\hat{x}= \operatorname*{\textbf{argmin}}_{x} \|y- Ax \|_2^2 + \lambda \sum_i \sum_{j \in \mathcal{N}_i} \| x[i]-x[j]\|^p
\end{equation}

\noindent where $p \leq 1$ and $\mathcal{N}_i$ is a neighborhood around the $i^{\text{th}}$ pixel. An iterative majorization-minimization algorithm is suggested for solving \eqref{eq:yang_inverse_problem}. Majorization of the regularization term will lead to the following quadratic surrogate problem:

\begin{equation} \label{eq:yang_update}
\hat{x}= \operatorname*{\textbf{argmin}}_{x} \|y- Ax \|_2^2 + \lambda \, x^T S x
\end{equation}

\noindent where $S$ is a sparse matrix representing the patch similarities. The algorithm alternates between minimization of \eqref{eq:yang_update} using a conjugate gradient descent method and update of the matrix $S$ from the new image estimate.

As we mentioned above, nonlocal patch similarities have been shown to be very useful for many image processing tasks. Because of space limitations, in this section we focused on image denoising and inverse problems, which are more relevant to CT. However, we should mention that in recent years, the idea of exploiting nonlocal patch similarities has been applied to many image processing tasks and this is currently a very active area of research. Some examples of these applications include image enhancements \cite{buades2006}, deblurring \cite{kindermann2005}, inpainting \cite{guillemot2014}, and super-resolution \cite{protter2009}.

\section{Other patch-based methods}
\label{sec:other_patch_based}

The large number and diversity of patch-based image processing algorithms that have been developed in the past ten years makes it impossible to review all of them here. Nonetheless, most of these algorithms are based on sparse representation of patches in learned dictionaries (Section \ref{sec:DL}) and/or exploiting nonlocal patch similarities (Section \ref{sec:NLM}). In this section, we try to provide a broad overview of some of the extensions of these ideas and other patch-based methods.

To begin with, it is natural to combine the two ideas of learned dictionaries and non-local filtering to enjoy the benefits of both methods. Research in this direction has proven to be very fruitful. The first algorithm to explicitly follow this approach was ``the non-local sparse model" proposed in \cite{mairal2009}. This method collects similar patches of the image, as in NLM denoising. However, unlike NLM that performs a weighted averaging, the non-local sparse model uses sparse coding of similar patches in a learned dictionary. The basic assumption in the non-local sparse model is that similar patches should use similar dictionary atoms in their representation. Therefore, simultaneous sparse coding techniques (e.g., \cite{tropp2006, tropp2006b}) are applied on groups of similar patches.

The idea of combining the benefits of non-local patch similarities and of learned dictionaries has been explored by many studies in the recent years \cite{chatterjee2009, dong2011, dong2011b, deledalle2011, yang2012b, tasdizen2009}. Most of these methods have reported state-of-the-art results. Although the details of these algorithms are different, the main ideas can be simply explained in terms of the non-local patch similarities and sparse representation in learned dictionaries. The K-LLD algorithm \cite{chatterjee2009}, for instance, uses steering kernel regression method to find structurally similar patches and then uses PCA to learn a suitable dictionary for each set of similar patches. The Adaptive Sparse Domain Selection (ASDS) algorithm \cite{dong2011}, on the other hand, clusters the training patches and learns a sub-dictionary for each cluster using PCA. For a new patch, then, ASDS selects the most relevant sub-dictionary for sparse coding of that patch. The idea of using PCA for building the dictionaries in these methods has received great attention because the learned dictionaries will be orthonogonal. In \cite{deledalle2011}, global, local, and hierarchical implementations of PCA dictionaries were studied. It was found the local-PCA (i.e., PCA applied on patches selected from a sliding window) led to the best results.

A very successful patch-based image denoising algorithm, that has similarities with the non-local sparse model, is the BM3D algorithm \cite{dabov2007}. Even though BM3D was proposed  in 2007, it is still regarded as the state-of-the-art image denoising algorithm. Similar to the non-local sparse model, BM3D collects similar patches and filters them jointly. However, unlike the non-local sparse model, it uses orthogonal DCT dictionaries instead of learned overcomplete dictionaries. Moreover, BM3D works in two steps. First, patch-matching and filtering is performed on the original noisy image to obtain an intermediate denoised image. Then, a new round of denoising is performed. This time, the intermediate image is used for finding similar patches. The algorithm includes other components such as Wiener filtering and weighted averaging \cite{dabov2007}. Further improvements to the original BM3D algorithm and an extension to 3D images (called the BM4D algorithm) have also been proposed \cite{dabov2009, maggioni2013}.

\section{Patch-based methods for Poisson noise}
\label{sec:poisson-patch}

In this section, we focus on the patch-based methods for the case when the noise follows a Poisson distribution. The reason for devoting a section to this topic is that, as we explained in Section \ref{sec:CT_Imaging_Model}, the noise in CT projection measurements has a complex distribution that can be best approximated as a Poisson noise or, after log-transformationm, as a Gaussian noise with signal-dependent variance \cite{macovski1983, wang2008}. In any case, application of the patch-based image processing methods to the projection measurements in CT requires careful consideration of the complex noise distribution. Unfortunately, most of the patch-based image processing methods, including all algorithms that we have described so far in this chapter, have been proposed for Gaussian noise. Moreover, most of these algorithms (with the exception of fully-Bayesian methods described in Section \ref{sec:DL_learning}) assume that the Gaussian noise has a uniform variance. Comparatively, the research on patch-based methods for the case of Poisson noise has been very limited and most of these limited works have been published very recently.

An important first obstacle facing the application of patch-based methods to the case of Poisson noise is the choice of an appropriate patch similarity measure. Methods that depend on nonlocal patch similarities need a patch similarity measure to find similar patches. Likewise, when we use sparse representation of the patches in a learned dictionary we often need a patch similarity measure. This is needed, for example, for finding the sparse representation of the patch in the dictionary using greedy methods. When the noise has a Gaussian distribution, the standard choice is the Euclidean distance, which has a sound theoretical justification and is easy to use.

For the non-Gaussian noise distributions, one straight-forward approach is to apply a so-called variance-stabilization transform so that the noise becomes close to Gaussian and then use the Euclidean distance. For the Poisson noise, the commonly-used transforms include the Anscombe transform \cite{anscombe1948} and the Haar-Fisz transform \cite{fryzlewicz2004}. If one wants to avoid these transforms and work with the original patches that are contaminated with Poisson noise, the proper choice of patch similarity measure is less obvious. Over the years, many criteria have been suggested for measuring the similarity between patches contaminated with Poisson noise \cite{alter2006}. For the case of low-count Poisson measurements, one study has suggested that the earth mover's distance (EMD) is a good measure of distance between patches \cite{giryes2013}. It has been suggested that EMD can be approximated by passing the patches with Poisson noise through a Gaussian filter and then applying the Euclidean distance \cite{giryes2013}. One study compared several different patch distance measures for Poisson noise through extensive numerical experiments \cite{deledalle2011b}. It was found that the generalized likelihood ratio (GLR) was the best similarity criterion in terms of the trade-off between the probability of detection and false alarm \cite{deledalle2011b}. GLR has many desirable theoretical properties that make it very appealing as a patch distance measure \cite{deledalle2012}. For the Poisson noise, this ratio is given by the following Equation:

\begin{equation} \label{eq:GLR_Poisson}
\mathcal{L}_G(x_1,x_2)= \frac{(x_1+x_2)^{x_1+x_2}}{2^{x_1+x_2} x_1^{x_1} x_2^{x_2}}
\end{equation}

Given two noisy patches $x_1[i]$ and $x_2[i]$, where $i \in \omega$, and assuming that the noise in pixels is independent, this gives the following similarity measure between the two patches:

\begin{equation} \label{eq:GLR_Poisson_patch}
\begin{aligned}
\mathcal{S}(x_1,x_2)= \sum_{i \in \omega} & (x_1[i]+x_2[i]) \log (x_1[i]+x_2[i]) \\
& - (x_1[i]) \log (x_1[i]) - (x_2[i]) \log (x_2[i])\\ & -(x_1[i]+x_2[i]) \log 2  \\
\end{aligned}
\end{equation}

In \cite{deledalle2012}, the GLR-based patch similarity criterion was also compared with six other criteria for non-local patch-based denoising of images with Poisson noise. It was found that using GLR led to the best denoising result when the noise is strong \cite{deledalle2012}. When the noise was not strong, the results showed that it was better to use a variance-stabilization transform to convert the Poisson noise into Gaussian noise and then to use the Euclidean distance. The algorithm used in \cite{deledalle2012} for non-local filtering is as follows:

\begin{equation} \label{eq:nlm_glr}
\widehat{x_0}(i)= \sum_{j=1}^N \frac{\mathcal{L}_G(x[j]-x[i])^{1/h}}{\sum_{j=1}^N \mathcal{L}_G(x[j]-x[i])^{1/h}} x(j)
\end{equation}

\noindent This algorithm includes the parameter $h$ instead of the kernel bandwidth $a$ in Equation \eqref{eq:nlm}.

Another nonlocal patch-based denoising algorithm for Poisson noise was suggested in \cite{deledalle2010}. A main feature of this algorithm is that the patch similarity weights are computed from the original noisy image as well as from a pre-filtered image:

\begin{equation} \label{eq:poisson_nl_means}
\widehat{x_0}(i)= \sum_{j=1}^N \frac{w_{i,j}}{\sum_{j=1}^N w_{i,j}} x(j)   \hspace{0.5cm}   \textbf{where:}   \hspace{0.5cm}   w_{i,j}= \exp \bigg( - \frac{u_{i,j}}{\alpha} - \frac{v_{i,j}}{\beta} \bigg)
\end{equation}

\noindent where $u_{i,j}$ are computed from the noisy image using a likelihood ratio principle and $v_{i,j}$ are computed from a pre-estimate of the true image using the symmetric Kullback-Leibler divergence. It is shown that the optimal values for the parameters $\alpha$ and $\beta$ can be computed and that this algorithm can achieve state of the art denoising results.

The patch-similarity measure in \eqref{eq:GLR_Poisson_patch} was used to develop a k-medoids denoising algorithm in \cite{chainais2012}. The k-medoids algorithm is similar to k-means algorithms. They are different in that k-means uses the centroid of each cluster as the representative of that cluster, whereas the k-medoids algorithm uses data points (i.e., examples) as the representative of the cluster. Moreover, k-medoids can work with any distance measure, not necessarily the Euclidean distance. It was shown in \cite{chainais2012} that the k-medoids algorithm achieved very good Poisson denoising results, outperforming the nonlocal Poisson denoising method of \cite{deledalle2012} in some tests. The k-medoids algorithm is in fact a special case of the dictionary learning approach. The difference with the dictionary-learning approach is that in the k-medoids algorithm only one atom participates in the representation of each patch.

The reason why the study in \cite{chainais2012} limited itself to using only one atom for representation of each patch was the difficulties in sparse coding under the Poisson noise. Suppose that $x_0[i]$ is the $i^{\text{th}}$ patch of the true underlying image and $x[i]$ is the measured patch under Poisson noise. If we wish to recover $x_0[i]$ from $x[i]$ via sparse representation in a dictionary $D$, we need to solve a problem that has the following form \cite{dupe2013}:

\begin{equation} \label{eq:sparse_coding_Poisson}
\widehat{\gamma_i} = \operatorname*{\textbf{argmin}}_{\gamma \text{ s.t. } \| \gamma \|_0 \leq T} \textbf{1}^T D \gamma_i - x[i]^T \log(D\gamma_i) \hspace{0.5cm}    \textbf{subject to:} \hspace{0.5cm} D\gamma_i >0
\end{equation}

\noindent Having found $\widehat{\gamma_i}$, we will have: $\widehat{x_0}[i]= D \widehat{\gamma_i}$. The difficulties of solving this problem have been discussed in \cite{dupe2013, giryes2013} and greedy sparse coding algorithms have been proposed for solving this problem. The author of \cite{dupe2013} then apply their proposed algorithm for denoising of images with Poisson noise. Even though they use a wavelet basis for $D$, they achieve impressive results.

A true dictionary learning-based denoising algorithm for images with Poisson noise was suggested in \cite{giryes2013}. In that study, a global dictionary is learned from a set of training data. Then, for a given noisy image to be denoised, the algorithm first clusters similar patches. All patches in a cluster are denoised together via simultaneous sparse representation in $D$. This means that patches that are clustered together are forced to share similar dictionary atoms in their representation. Experiments showed that this method was comparable with or better than competing methods. A slightly similar approach that also combines the ideas of learned dictionaries and non-local filtering is proposed in \cite{salmon2012, salmon2014}. In this approach, k-means clustering is used to group similar image patches. A dictionary is learned for each cluster of similar patches using the Poisson-PCA algorithm \cite{collins2002, singh2008}. For solving the Poisson-PCA problem, which is also known as exponential-PCA, the authors use the Newton's method. This algorithm showed good performance under low-count Poisson noise.

\section{Total variation (TV)}
\label{sec:tv_basics}

Total variation (TV), which was fist proposed in \cite{rudin1992} for image denoising and reconstruction, has become one of the most widely used regularization functions in image processing. For an image $x(u)$ defined on $\Omega \subset \mathbb{R}^2$ it is defined as \cite{chambolle2010}:

\begin{equation} \label{eq:tv_basic_def}
\begin{aligned}
J(x)= \text{sup} & \Bigg\{ \int_{\Omega} x(u) \text{div} \Phi(u) du: \\ & \hspace{1cm} \Phi \in C_c^1(\Omega, \mathbb{R}^2), |\Phi(u)| \leq 1 \, \forall u \in \Omega  \Bigg\}
\end{aligned}
\end{equation}

\noindent For a smooth image $x \in C^1(\Omega)$ it takes the form:

\begin{equation} \label{eq:tv_continuous}
J(x)= \int_{\Omega} |\nabla x | du
\end{equation}

Many different discretizations have been proposed. Suppose $x \in \mathbb{R}^{N \times N}$ is a 2D image. A common discretization is \cite{chambolle2004}:

\begin{equation} \label{eq:tv_discrete}
J(x)= \sum_{1 \leq i,j \leq N} |(\nabla x)_{i,j}|
\end{equation}

\noindent where

\begin{equation} \label{eq:tv_discrete2}
\begin{aligned}
(\nabla x)_{i,j}=& \big((\nabla x)^1_{i,j},(\nabla x)^2_{i,j} \big) \\
& \hspace{1cm} (\nabla x)^1_{i,j}= \left\{ \begin{array}{ll}
x_{i+1,j}-x_{i,j} & \text{if } i<N \\
0 & \text{if } i=N\\
\end{array}   \right. \\
& \hspace{1cm} (\nabla x)^2_{i,j}= \left\{ \begin{array}{ll}
x_{i,j+1}-x_{i,j} & \text{if } j<N \\
0 & \text{if } j=N\\
\end{array}   \right. \\
\end{aligned}
\end{equation}

\noindent and for $z=(z_1,z_2) \in \mathbb{R}^2$, $|z|= \sqrt{z_1^2+z_2^2}$.

Suppose that we obtain measurements $y= Ax+w$, where, as before, $A$ is some operation or transformation such as blurring, sampling, or forward projection in CT and $w$ is additive Gaussian noise with uniform variance. The maximum a posteriori estimate of $x$ with a total variation prior $P(x) \sim e^{-J(x)}$ is obtained as:

\begin{equation} \label{eq:tv_deconvolution}
x_{\text{MAP}}= \operatorname*{\text{argmin}}_{x} \|Ax-y \|_2^2 + J(x)
\end{equation}

A special case of this problem is the denoising problem shown below, which corresponds to the case where $A$ is the identity matrix.

\begin{equation} \label{eq:ROF_tv_denoising}
x_{\text{MAP}}= \operatorname*{\text{argmin}}_{x} \|x-y \|_2^2 + \int_{\Omega} |\nabla x | du
\end{equation}

\noindent which is usually referred to as the Rudin-Osher-Fatemi (ROF) model for image denoising.

The main properties of TV include convexity, lower semi-continuity, and homogeneity \cite{chambolle2010}. Many different algorithms have been suggested for solving this problem. Examples of the optimization approaches that are used to solve this problem include primal-dual methods \cite{krishnan2007, vogel2002}, second-order cone programming \cite{goldfarb2005}, dual formulations \cite{chambolle2004, zhu2008}, split Bregman methods \cite{goldstein2009, getreuer2012}, and accelerated proximal gradient methods \cite{beck2009b, nesterov2007}.

In general, TV is a good model for recovering blocky images, i.e., images that consist of piecewise-constant features with sharp edges \cite{vogel2002}. Many studies have used TV to successfully accomplish various image processing tasks, including denoising \cite{chambolle2004}, deblurring \cite{wu2010}, inpainting \cite{papafitsoros2013}, restoration \cite{bioucas-dias2007}, and reconstruction \cite{becker2009}. However, on images with fine texture and ramp-like features, this model usually performs poorly \cite{gilboa2006b}. Therefore, many studies have tried to improve or modify this model so that it can be useful for more complicated images. Some of the research directions include employing higher-order differentials \cite{bredies2010, benning2012, bergounioux2010}, locally adaptive formulations that try to identify the type of local image features and adjust the action of the algorithm accordingly \cite{guo2009, chen2010, dong2011c, grasmair2009}, and combining TV with other regularizations in order to improve its performance \cite{grasmair2010, lysaker2006}.

\section{Published research on sparsity-based methods in CT}
\label{sec:review-ct}

This section reviews some of the published research on the application of the sparsity-based models and algorithms described so far in this chapter in CT. We divide these applications into three categories: 1) pre-processing methods, which aim at restoring or denoising of the projection measurements, 2) iterative reconstruction methods, and 3) post-processing methods, whose goal it to enhance, restore, denoise, or otherwise improve the quality of the reconstructed image.

\subsection{Pre-processing methods}

Compared with iterative reconstruction methods and post-processing methods, pre-processing methods account for a much smaller share of the published studies on sparsity-based algorithms for CT \cite{karimi2016sinoprojection,karimi2015angular,karimi2016joint}. There are two main reasons behind this. The first reason is that the pre-processing methods for CT, in general, face certain difficulties. For example, it is well-known that sharp image features are smoothed in the projection domain. Therefore, preservation of sharp image features and fine details is more challenging when working in the projection domain. Moreover, many commercial scanners do not allow access to the raw projection data. Therefore, it is more difficult to validate the pre-processing algorithms and apply them in clinical settings. The second reason is that the great majority of the sparsity-based image processing algorithms have been proposed with the assumption of additive Gaussian noise with uniform variance. As we described in Section \ref{sec:poisson-patch}, research on patch-based methods for the case of Poisson noise has been much more limited in extent and the algorithms that have been proposed for Poisson noise are very recent and have not yet been absorbed by researchers working on CT.

A patch-based sinogram denoising algorithm was proposed in \cite{shtok2013}. A fixed DCT dictionary was used for representation of the sinogram patches. However, the shrinkage rule used for denoising was learned from training data. The denoised projections were then used to reconstruct the image using an FBP method. A patch-based processing using learned shrinkage functions is then applied on the reconstructed image. The results of the study showed that this rather simple algorithm outperformed some of the well-known iterative CT reconstruction algorithms.

Use of learned dictionaries for inpainting (i.e., upsampling) of the CT projection measurements has also been proposed \cite{li2014}. The goal of sinogram upsampling is to reduce the x-ray dose used for imaging by measuring only a fraction of the projections directly and estimating the unobserved projections with upsampling. The assumption used in this algorithm was that patches extracted from the projections admit a sparse representation in a dictionary that could be learned from a set of training sinograms. The approach followed by this study was very similar to the general inpaining approach that we explained in Section \ref{sec:DL_applications}. The results of the study showed that dictionary-based upsampling of the projections substantially improved the quality of the images reconstructed with FBP, outperforming more traditional sinogram interpolation methods based on splines.

As we mentioned above, a challenge for all sinogram denoising/restoration algorithms is preservation of fine image detail. The algorithm presented in \cite{shtok2011} has proposed an interesting idea to address this issue. In fact, this study contains several interesting ideas. One of these ideas is that in learning a dictionary for sparse representation of sinogram patches, not only the sinogram-domain error but also the error in the image domain is considered. Specifically, first a dictionary ($D_1$) is learned considering only the error in the sinogram domain. Let us denote the CT image and its sinogram with $x$ and $y$, respectively. Then $D_1$ is found by solving:

\begin{equation} \label{eq:shtok_d1}
\{D_1, \hat{\Gamma} \}= \operatorname*{\textbf{argmin}}_{D, \Gamma} \|\Gamma \|_0 \hspace{0.5cm} \textbf{subject to:} \hspace{0.5cm} \| D \Gamma_i - R_i y \|_2^2 \leq C \sigma_i \;\;\; \forall \, i
\end{equation}

This optimization to find $D_1$ is carried out using the K-SVD algorithm that we described in Section \ref{sec:DL}. The only difference here is that signal-dependent nature of noise, $\sigma_i$, should be taken into account in the sparse coding step ($C$ is a tuning parameter). This dictionary is then further optimized by minimizing the reconstruction error in the image domain:

\begin{equation} \label{eq:shtok_d2}
D_2= \operatorname*{\textbf{argmin}}_{D} \Bigg\|\mathcal{FBP}\bigg(\sum_i (R_i^TR_i)^{-1} \sum_i R_i^T D \hat{\Gamma}\bigg) - x \Bigg\|_{Q,2}^2
\end{equation}

\noindent where we have used $\mathcal{FBP}$ to denote the CT reconstruction algorithm (here, filtered back-projection). Note that the $\hat{\Gamma}$ in the above optimization problem is that found by solving \eqref{eq:shtok_d1}. In other words, for finding $D_2$ we keep the sparse representations fixed and find a dictionary that leads to a better reconstruction of the image, $x$. The notation $\| . \|_{Q,2}$ denotes a weighted $\ell_2$ norm. It is suggested that the weights $Q$ are chosen such that more weight is given to low-contrast features \cite{shtok2011}.

The $x$ and $y$ in the above equations denote the ``training data", which includes a set of high-quality images and their projections. In fact, instead of only one image, a large number of images can be used for better training. Now, suppose that we are given noisy projections of a new object/patient, which we denote with $y_{\text{noisy}}$. It is suggested to denoise $y_{\text{noisy}}$ in two steps. First, sparse representations of patches of $y_{\text{noisy}}$ in $D_1$ are obtained. Denoting this with $\hat{\Gamma}$, the final denoised sinogram is obtained as the solution of the following problem which uses $D_2$:

\begin{equation} \label{eq:shtok_denoising}
y_{\text{denoised}}= \operatorname*{\textbf{argmin}}_{y} \lambda \| y- y_{\text{noisy}} \|_W^2 + \sum_i \| D_2 \hat{\Gamma}_i - R_i y \|
\end{equation}

\noindent where $W$ are weights to account for the signal-dependent nature of the noise. This problem has a simple solution similar to Equation \eqref{eq:KSVD_denoising_update}. Experimental results for 2D CT have shown promising results \cite{shtok2011}. Finally, TV-based sinogram denoising algorithms were propsoed in \cite{karimi2016localtv,karimi2016bregman}.

\subsection{Iterative reconstruction methods}

In recent years, several iterative image reconstruction algorithms involving regularizations in terms of image patches have been proposed for CT. In general, these algorithms have reported very promising results. However, a convincing comparison of these algorithms with other classes of iterative reconstruction algorithms such as those based on TV or other edge-preserving regularizations \cite{karimi2015weighted,karimi2017stochastic,karimi2017sparseview,karimi2016hybrid} is still lacking. In this section, we review some of the iterative CT reconstruction algorithms that use patch-based or TV regularization.

A typical example of dictionary-based CT reconstruction algorithms is the algorithm proposed in \cite{xu2012}. That paper suggested recovering the image as a solution of the following optimization problem:

\begin{equation} \label{eq:dictionary_xu}
\operatorname*{\textbf{minimize}}_{x,D,\Gamma} \sum_i w_i ([Ax]_i-y_i))^2 + \lambda \left( \sum_k || R_k x - D \Gamma_k ||_2^2 + \nu_k ||\Gamma_k||_0  \right)
\end{equation}

In the above problem, $A$ is the projection matrix \cite{karimi2015computational} and $w_i$s are noise-dependent weights. The first term in the objective function encourages measurement consistency. The remaining terms constitute the regularization, which are very similar to the terms in the formulation of the basic dictionary learning problem in \eqref{eq:DL_olshausen}. In \eqref{eq:dictionary_xu}, the dictionary is learned from the image itself. The authors of \cite{xu2012} solved this problem by alternating minimization with respect to the three variables. Minimization with respect to $x$ is carried out using the separable paraboloid surrogate method suggested in \cite{elbakri2002}. The problem with this approach, however, is that it requires access to the individual elements of the projection matrix. Although this is a simple requirement for 2D CT, for large 3D CT, this can be a major problem because with efficient implementations of forward and back-projection operations it is not convenient to access individual matrix elements \cite{long2010, karimi2015computational,karimi2016thesis}. Minimization with respect to $D$ and $\Gamma$ is performed using the K-SVD and OMP algorithms, respectively. Alternatively, the dictionary can be learned in advance from a set of training images. This will remove $D$ from the list of the optimization variables in \eqref{eq:dictionary_xu}, substantially simplifying the problem. Both approaches are indeed presented in \cite{xu2012}. Experiments showed that both of these approaches led to very good reconstructions, outperforming a TV-based algorithm. 

Formulations very similar to the one described above were shown to be superior to TV-based reconstruction and other standard iterative reconstruction algorithms in electron tomography \cite{alafeef2014, liu2014}. Another study first learned a dictionary from training images, but for image reconstruction step did not include the sparsity term in the objective function \cite{Etter2011}. In other words, only the first two terms in the objective function in Equation \eqref{eq:dictionary_xu} were considered. A gradient descent approach was used to solve the problem. That study found superior reconstructions with learned dictionaries compared with a DCT basis.

One study used an optimization approach similar to the one described above, but used box-splines for image representation \cite{sakhaee2014}. In other words, instead of native pixel representation of the image, box spline were used as the basis functions in the image domain. The unknown image $x$ will have a representation of the form $x= \sum_i c_i \phi_i$, where $\phi_i$ is the box spline centered on the $i^{\text{th}}$ pixel and $c_i$ is the value of attenuation coefficient for that pixel. The resulting optimization problem will have the following form:

\begin{equation} \label{eq:sakhaee_image}
\operatorname*{\textbf{minimize}}_{c,\Gamma} \| H c-y \|_W^2 + \lambda \left( \sum_k || R_k c - D \Gamma_k ||_2^2 + \nu_k ||\Gamma_k||_0  \right)
\end{equation}

\noindent In the above problem, $H$ is the forward model relating the image representation coefficients to the sinogram measurements, $y$. In other words, $H$ is simply the equivalent of the projection matrix $A$. The rest of the objective function is the same as that in Equation \eqref{eq:dictionary_xu}. Once the representation coefficients, $c$, are found by solving \eqref{eq:sakhaee_image}, the image is reconstructed simply as $x= \sum_i c_i \phi_i$. The results of the study showed that this dictionary-based algorithm achieved much better reconstructions than a wavelet-based reconstruction algorithm.

The dual-dictionary methods proposed in \cite{lu2012, zhao2012} rely on two dictionaries. One of the dictionaries ($D_l$) is composed of patches from CT images reconstructed from a small number of projection views, while the second dictionary ($D_h$) contains the corresponding patches from a high-quality image. The atoms of the two dictionaries are in one-to-one correspondence. The strategy here is to first find the sparse code of the patches of the image to be reconstructed in $D_l$ and then to recover a good estimate of the patch by multiplying this sparse code with $D_h$. The dictionaries are not learned here, but they are built by sampling a large number of patches from few-view and high-quality training images. This approach has been reported to achieve better results than TV-based reconstruction algorithms \cite{lu2012}.

A different dictionary-based reconstruction algorithm was suggested in \cite{soltani2015}. In this algorithm, first a dictionary ($D$) is learned by solving a problem of the following form:

\begin{equation} \label{eq:DL_soltani}
\operatorname*{\textbf{minimize}}_{D,\Gamma} \|X - D \Gamma \|_F^2 + \lambda \|\Gamma \|_{1} \hspace{0.3cm} \text{\textbf{subject to:}} \hspace{0.3cm} D \in \mathcal{D} \hspace{0.2cm} \& \hspace{0.2cm} \Gamma \in \mathbb{R}_+
\end{equation}

\noindent where $\mathcal{D}$ can be an $\ell_2$ ball and $\mathbb{R}_+$ is the non-negative orthant of the proper size. The above problem is solved using the Alternating Direction Method of Multipliers (ADMM) to find the dictionary. It is reported that learning the dictionary with ADMM is computationally very efficient and largely independent of the initialization. The learned dictionary is then used to regularize the reconstruction algorithm by requiring that the patches of the reconstructed image have a sparse representation in the dictionary. However, unlike most other dictionary-based algorithms, overlapping patches are not used. Instead, a novel regularization term is introduced to avoid the blocking artifacts at the patch borders. Specifically, the optimization problem to recover the image from projection measurements $y$ has this form:

\begin{equation} \label{eq:Reconstruct_soltani}
\operatorname*{\textbf{minimize}}_{x_{\Gamma}} \|A x_{\Gamma} - y \|_2^2 + \lambda \|\Gamma \|_{1} + \mu \|L x_{\Gamma} \|_2^2
\end{equation}

\noindent where, to simplify the notation, we have used $x_{\Gamma}$ to emphasize that the reconstructed image depends on the sparse representation matrix, $\Gamma$. The matrix $L$ is a matrix that computes the directional derivatives across the patch boundaries. Therefore, the role of the last term in the objective function is to penalize large jumps at the patch boundaries, thereby suppressing blocking artifacts that arise when non-overlapping patches are used. Comparison with TV-based reconstruction showed that this dictionary-based reconstruction algorithm reconstructed mush better images, preserving fine textural detail that are smeared by TV-based reconstruction. Overall, the algorithm proposed in that paper contains several interesting ideas that can be useful for designing dictionary-based reconstruction algorithms for CT. A later paper studied the sensitivity of this algorithm to such factors as the scale and rotation of features in the training data \cite{soltani2015b}.

An iterative reconstruction algorithm that combines sparse representation of image patches with sinogram smoothing was proposed in \cite{stojanovic2012}. The image is reconstructed as a solution of the following optimization problem:

\begin{equation} \label{eq:dictionary_stojanovic}
\begin{aligned}
\operatorname*{\textbf{minimize}}_{x,y,\Gamma} & \|y- \bar{y}\|+ \alpha \, y^T W y + \beta \| Ax-y\|_2^2 \\ & + \lambda \left( \sum_k || R_k x - D \Gamma_k ||_2^2 + \nu_k ||\Gamma_k||_0  \right)
\end{aligned}
\end{equation}

The first two terms, where $\bar{y}$ is the measured noisy sinogram, represent the sinogram Markov random
field model \cite{li2004, wang2006}. The remaining terms are similar to those we encountered above. As usual, the above problem is suggested to be solved using a block-coordinate minimization, where the minimization with respect to the image $x$ is suggested to be carried out using a conjugate gradients method. That study also suggests interesting variations of the objective function in \eqref{eq:dictionary_stojanovic}, but the experimental evaluations that are presented are very limited.

As the last example of dictionary-based iterative reconstruction algorithms, we should also mention the method based on sparsifying transforms that was proposed in \cite{pfister2013, pfister2014}. Sparsifying transforms are variations of the analysis model for sparsity \cite{ravishankar2013, wen2015}. In the analysis model, instead of the relation $x= D \gamma$ that we have discussed so far in this chapter, we have $Dx= \gamma$. In other words, $D$ acts as an operator on the signal (e.g., the image patch) to find the representation coefficients, $\gamma$. In \cite{pfister2013, pfister2014}, it is suggested that the unknown CT image be recovered as the solution of the following optimization problem:

\begin{equation} \label{eq:DL_analysis}
\begin{aligned}
& \operatorname*{\textbf{minimize}}_{x,D,\Gamma} \sum_i \| D R_ix - \Gamma_i \|_2^2 + \lambda \|\Gamma \|_{1} + \alpha H(D) \hspace{0.5cm} \\ & \text{\textbf{subject to:}} \hspace{0.5cm} \|Ax-y \|_W^2 \leq \epsilon
\end{aligned}
\end{equation}

\noindent where $H(D)$ is a regularization on the dictionary $D$, and $W$ represents the weights introduced to account for the signal-dependent noise variance. The results of that study showed that this approach led to results that were comparable with iterative reconstruction with synthesis formulation and TV-based regularization, while also being slightly faster.

In recent years, there has also been a growing attention to the potential of regularization in terms of non-local patch priors for iterative CT reconstruction. In \cite{lou2010}, it was suggested to recover the CT image as a solution of the following optimization problem:

\begin{equation} \label{eq:lou_inverse_problem}
\hat{x}= \operatorname*{\textbf{argmin}}_{x} \|y- Ax \|_2^2 + \lambda J_{\text{NL}}(x)
\end{equation}

\noindent where $J_{\text{NL}}(x)$ is the regularization in terms of patch similarities. Two different forms were suggested for $J_{\text{NL}}(x)$:

\begin{equation} \label{eq:lou_regularization}
\begin{aligned}
J_{\text{NL/TV}}(x)= & \sum_i \sum_{j \in \mathcal{N}_i} \| \sqrt{w_{i,j}} (x(i)-x(j))\|_2 \\
J_{\text{NL/}H\textsuperscript{1}}(x)= & \sum_i \sum_{j \in \mathcal{N}_i} w_{i,j} \|x(i)-x(j)\|^2_2
\end{aligned}
\end{equation}

\noindent where $w_{i,j}$ are the patch-based similarity weights. For the $i^{\text{th}}$ pixel, they are computed from all pixels $j$ in a window around $i$ using:

\begin{equation} \label{eq:lou_weights}
w_{i,j}= \exp{\bigg(-\frac{\|x[i]-x[j] \|}{h^2} \bigg)}
\end{equation}

\noindent It is suggested that these weights be computed from a FBP-reconstructed image and that the filter parameter $h$ be chosen based on the local estimate of noise variance. The local noise variance is estimated from the wavelet coefficients of the finest wavelet subband ($v_i$) according to \cite{donoho1994}:

\begin{equation} \label{eq:lou_noise_estimate}
h= \frac{\text{median}(|v_i|)}{0.6745}
\end{equation}

The authors of \cite{lou2010} solved the problem \eqref{eq:lou_inverse_problem} with either of the regularization functions in \eqref{eq:lou_regularization} using a simple gradient descent and found that the recovered CT image had a better visual and objective quality than a standard TV-based iterative reconstruction algorithm.

As simple iterative algorithm that alternates between projections onto convex sets (POCS) to improve measurements consistency and an NLM-type restoration has been proposed in \cite{huang2011} . That algorithm was shown to perform better than a TV-based algorithm but no comparison with the state of the art methods was performed. Another study developed a NLM-type regularization for perfusion CT that relies on a high-quality prior image \cite{ma2012}. The proposed regularization function, shown in the following equation, is in terms of the similarity between patches of the unknown image to be reconstructed from a low-dose scan ($x$) and the patches of the prior image ($x_p$).

\begin{equation} \label{eq:nlm_MA}
J(x)= \| x- \bar{x}\|_q^q \hspace{0.5cm} \textbf{where} \hspace{0.5cm} \bar{x}(i)= \sum_{j \in \mathcal{N}_i} w_{i,j} x_p(j)
\end{equation}
The authors suggest $q=1.2$. A steepest-descent approach is used to approximately solve this problem. A similar, but more general, algorithm that does not require a prior image was proposed in \cite{zhang2014}. The formulation is the same as the above, the main difference being that the weights in the NLM formulation are computed from the image itself. A Gauss-Seidel approach is used to solve the resulting problem. Both of the above NLM-type regularization methods are reported to result in better reconstructions than more conventional regularizations such as Gaussian Markov random field. 

Non-local patch-based regularization was also used for the new technique of equally-sloped tomography (EST, \cite{miao2005}) and was shown to improve the quality of the reconstructed image both from small or large number of projections \cite{fahimian2010}. Nonlocal patch-based regularization substantially improved the CNR, SNR, and spatial resolution of the images reconstructed from 60, 90, and 360 projections in that study.

Patch-based iterative reconstruction algorithms have also been proposed for dynamic CT. In dynamic CT, several successive images of the the same patient are reconstructed. Therefore, there is abundant temporal correlation (i.e., correlation between successive images) in addition to the spatial correlation within each of the images in the sequence. There have been several studies in recent years that have aimed at exploiting these correlations in terms of patch/block similarities. In general, these studies have reported very promising results.

A reconstruction algorithm with nonlocal patch-based regularization was proposed for dynamic CT in \cite{kazantsev2015}. The proposed regularizer for the $k^{\text{th}}$ frame of the image is as follows:

\begin{equation} \label{eq:nlm_kazantsev}
\begin{aligned}
J(x_k)=  &\sum_i \sum_{j \in \mathcal{N}_i} G_a (x_k[i]-x_k[j]) |x_k(i)-x_k(j)|^2 \\
 & + \sum_i \sum_{l \in \{1,2, ..., K  \} \setminus k} \sum_{j \in \Delta_i} G_a (x_l[i]-x_l[j]) |x_l(i)-x_l(j)|^2
\end{aligned}
\end{equation}

\noindent where, as before, $x(i)$ and $x[i]$ denote the $i^{\text{th}}$ image pixel and the patch centered on that pixel, respectively. $G_a(.)$ is a Gaussian kernel as in the standard NLM denoising. The first term is a spatial regularization in terms of the patches of the current image frame, $x_k$. In this term, $\mathcal{N}_i$ is a simple rectangular neighborhood around the $i^{\text{th}}$ pixel. The second term, where $K$ is the total number of frames, is a temporal patch-based regularization that involves patches from all other frames in the image sequence. In this term, $\Delta_i$ is a neighborhood whose spatial size if pre-fixed but whose temporal extension is found for each pixel such that the probability of finding patches with similar structural features (e.g., edges) is increased. This is done by dividing the temporal neighborhood into blocks and estimating the structural similarity of these blocks with the patch centered on the $i^{\text{th}}$ pixel. Only a small fraction of blocks that are most similar to $x[i]$ are included in $\Delta_i$. A similar approach was proposed in \cite{kazantsev2015b} for the case when a high-quality prior image is available. This high-quality prior image does not have to be a CT image and can be acquired in other imaging modalities. The results of experiments with simulated and real data show that this algorithm achieves very good reconstructions.

Temporal non-local-means (TNLM) algorithms were proposed in \cite{tian2011, jia2012}. These algorithms suggest recovering a set of successive CT images $\{x_k | \; k \in 1:K  \}$ by minimizing an optimization problem that includes (in addition to the measurement fidelity term) the following regularization:

\begin{equation} \label{eq:regularization_jia}
J\big(\{x_k\}\big) = \sum_{k=1}^K \sum_i \sum_j w_{i,j} \big( x_k(i)- x_{k+1}(j) \big)^2
\end{equation}

\noindent where, as usual, the weights are computed based on patch similarities:

\begin{equation} \label{eq:weights_jia}
w_{i,j}= \frac{1}{Z} \exp \bigg(- \frac{\|x[i]-x[j] \|^2}{2 h^2}   \bigg)
\end{equation}

An important choice in this algorithm is that only inter-image patch similarities are taken into account and not the intra-image patch similarities. The justification is that the proposed algorithm is for the case when each of the images in the sequence is reconstructed from a small number of projections and, hence, contains much streak artifacts. Therefore, using patches from the same image will amplify the streak artifacts, while using patches from neighboring images will suppress the artifacts. In addition to the iterative reconstruction algorithm, in \cite{jia2012} another very similar algorithm has been suggested that can also be classified as a post-processing algorithm. In this alternative scheme, each of the images in the sequence are first reconstructed from their correponding projections, and then they are post-processed using an optimization algorithm that includes the very same regularization function in \eqref{eq:regularization_jia}.

A tensor-based iterative reconstruction algorithm was proposed for dynamic CT in \cite{tan2015}. Tensor-based dictionaries are a relatively new type of dictionary that are gaining more popularity. As we have mentioned above, in image processing applications, image patches/blocks are vectorized and used as training/test signals. Tensor-based methods treat the image patches or blocks in their original form, i.e., without vectorizing them \cite{cichocki2015, caiafa2013}. Therefore, they are expected to better exploit the correlation between adjacent pixels. In \cite{tan2015}, a tensor-based algorithm was compared with a standard dictionary for dynamic CT reconstruction and it was found that the tensor-based dictionaries result in a slight improvement in the quality of the reconstructed image. 

Compared with the reconstruction algorithms that are based on learned dictionaries or nonlocal patch similarities, many more algorithms have used TV regularization terms. This is partly because TV-regularized cost functions are easier to handle using standard optimization algorithms, especially for large-scale 3D image reconstruction. Moreover, the CT community is more familiar with TV-based regularization because it has been used for CT reconstruction for a longer time. Many studies have formulated the reconstruction problem as a regularized least-squares minimization similar to \eqref{eq:tv_deconvolution}. Some of the optimization techniques that have been suggested for solving this problem include accelerated first-order methods \cite{choi2010, park2012, Jensen2011}, alternating direction method of
multipliers \cite{ramani2011}, and forward-backward
splitting algorithm \cite{jia2010}. Another very commonly used formulation for CT reconstruction is the constrained optimization formulation, where the image TV is minimized under measurement consistency constraints \cite{niu2012, han2011, ritschl2011}. Most published studies use an alternating algorithm for solving this problem, whereby at each iteration the image TV is reduced followed by a step that enforces the measurement consistency constraint. A simple (and probably inefficient \cite{chambolle2010}) method that has been adopted in many studies uses a steepest descent for TV minimization followed by projection onto convex sets for measurement consistency \cite{sidky2008}.

Several studies have combined the TV regularization with regularization in terms of a prior high-quality image in applications such as dynamic CT \cite{chen2008, bergner2010}, perfusion imaging \cite{nett2010}, and respiratory-gated CT \cite{leng2008, song2007}. In general, the existence of a high-quality prior image reduces the number of projection measurements required for reconstructing high-quality images from subsequent scans. Other variations of the standard TV regularization that have been successfully applied for CT reconstruction include non-convex TV \cite{chartrand2013, sidky2007} and higher-order TV \cite{yang2010b}.

In general, TV-based reconstruction methods have proven to be much better than  traditional CT reconstruction algorithms, particularly in reconstruction from few-view and noisy projection data. Therefore, many studies have concluded that TV-based reconstruction methods have a great potential for dose reduction in a wide range of CT applications \cite{bian2010, kim2012, jia2011b, tang2009}. However, there has been no satisfying comparison between TV and other edge-preserving or smoothness-promoting regularization functions that are very widely used in CT \cite{bouman1993, kim2015, wang2009b, delaney1998}.

\subsection{Post-processing methods}
\label{sec:CT_post_processing}

Many of the sparsity-based algorithms that have been proposed for CT fall into the category of post-processing methods \cite{karimi2016structured,karimi2016coupled}. This is partly because most of the sparsity-based algorithms that have been developed for CT are directly based on the sparsity-based methods that have been proposed for natural images. Because general sparsity-based image processing algorithms mostly include denoising and restoration algorithms, they are more easily extended as post-processing methods for CT. Moreover, some of the sparsity-based methods, particularly patch-based image processing methods, are very computationally expensive. Therefore, especially for large-scale 3D CT, it is easier to deploy them as one-shot post-processing algorithms than as part of an iterative reconstruction algorithm.

A large number of dictionary-based algorithms have been proposed for CT denoising. The basic denoising algorithm that we described in Section \ref{sec:DL_applications} was used for denoising of abdomen CT images in \cite{chen2013, chen2013c}, and head CT images in \cite{chen2014b} and showed promising results in all of these studies. Straightforward representation of image patches in a learned dictionary followed by weighted averaging resulted in effective suppression of noise and artifacts and a marked improvement in the visual and objective image quality.

Non-local means methods have also been applied for CT image denoising. An early example is \cite{kelm2009}. In that study, the authors investigated the effect of different parameters such as the patch size, smoothing strength, and the size of the search window around the current pixel to find similar patches. Among the findings of that study with lung and abdomen CT images was that one can choose the size of the search window for finding similar patches to be as small as $25 \times 25$ pixels and still achieve very impressive denoising results. However, this required careful tuning of the denoising parameter ($a$ in Equation \eqref{eq:nlm_practical}). Moreover, choosing a small search window also required reducing the patch size to ensure that for every pixel a sufficient number of similar patches are found in the search window. Otherwise, in certain areas such as around the edges, very little denoising is accomplished. Another study found that with a basic NLM denoising, the tube current setting can be reduced to one fifth of that in routine abdominal CT imaging without jeopardizing the image quality \cite{chen2011}.

An algorithm specially tailored to image-guided radiotherapy was proposed in \cite{yan2013}. Since in this scenario a patient is scanned multiple times, it was suggested that the first scan be performed with standard dose and later scans with much reduced dose. An NLM-type algorithm was suggested to reduce the noise in the low-dose images. The proposed algorithm denoised the low-dose images by finding similar patches in the image reconstructed from the standard-dose scan. Similarly, in CT perfusion imaging and angiography the same patient is scanned multiple times. A modified NLM algorithm was suggested for these imaging scenarios in \cite{ma2011}. The algorithm proposed in that study registers a standard-dose prior image to the low-does image at hand. The low-dose image is then denoised using a NLM algorithm where patches are extracted from the registered standard-dose image.

One study suggested adapting the strength of the NLM denoising based on the estimated local noise level \cite{li2014b}. That paper proposed a fast method for approximating the noise level in the reconstructed image and suggested choosing the bandwidth of the Gaussian kernel in the NLM denoising to be proportional to the estimated standard deviation of the noise. Evaluations showed that this algorithm effectively suppressed the noise without degrading the spatial resolution. Using speed-up techniques such as those in \cite{darbon2005b}, this algorithm was able to process large 3D images in a few minutes when implemented on GPU. 

Applying the nonlocal patch-based denoising methods in a spatially adaptive fashion has been proposed by many studies on natural images \cite{kervrann2006, kervrann2008}. For CT images, it is well known that the noise variance in the reconstructed image can vary significantly across the image. Therefore, estimating the local noise variance may improve the performance of the patch-based denoising methods. Another approach for estimating the local noise variance in the CT image was propose in \cite{bartuschat2009}. In this approach, which is much simpler than the method proposed in \cite{li2014b}, even and odd-numbered projections are used to reconstruct two images. Then, assuming the noise in the projections are uncorrelated, the local noise variance is estimated from the difference of the two images.

So far in this section, we have talked about algorithms that have been suggested primarily for removing the noise. However, CT images can also be marred by various types of artifacts that can significantly reduce their diagnostic value \cite{barrett2004}. Recently, a few patch-based algorithms have been proposed specifically for suppressing these artifacts. A dictionary-based algorithm for suppressing streak artifacts in CT images is proposed in \cite{chen2014}. The artifact-full image is first decomposed into its high-frequency bands in the horizontal, vertical, and diagonal directions. Sparse representation of patches of each of these bands are found in a set of three ``discriminative" dictionaries that include atoms specifically learned to represent artifacts and genuine image features. Artifacts are suppressed by simply setting to zero the large coefficients that correspond to the artifact atoms. The results of this study on artifact-full CT images are impressive.

A nonlocal patch-based artifact reduction method was suggested in \cite{xu2012b}. This method is tailored for suppressing the streak artifacts that arise when the number of projections used for image reconstruction is small and it relies on the existence of a high-quality prior image. The few-view image that is marred by artifacts is first registered to the high-quality reference image using a registration algorithm that uses the SIFT features \cite{lowe2004}. The registered reference image is then used to simulate an artifact-full few-view image. To remove the streak artifacts from the current image, its patches are matched with the simulated artifact-full image, but then the corresponding high-quality patches from the reference image are used to build the target image. This algorithm is further extended in \cite{xu2013} to be used when a prior scan from the same patient is not available but a rich database of scans from a large number of patients exists. The results of both of these studies on real CT images of human head and lung are very good. Both of the methods substantially reduced the streaking artifacts in images reconstructed from less than 100 projections.

A major challenge facing the application of patch-based algorithms for large 3D CT images is the issue of the computational time. Although we discuss this challenge here under the post-processing methods, they apply equally to pre-processing methods and are indeed even more relevant to iterative reconstruction algorithms. Of course, one obvious approach to reducing the computational load is to work with 2D patches, instead of 3D blocks. However, this will likely hurt the algorithm performance because the voxel correlations in the 3rd dimension are not exploited. Three studies have reported that compared with 2D denoising, 3D denoising of CT images leads to an improvement in PSNR of approximately $1$ to $4 \, \text{dB}$ \cite{rubinstein2010b, li2012, li2012c}. Another study used 2D patches to denoise the slices in 3D CT images but they used patches from neighboring slices in addition to patches from the same slice \cite{kelm2009}. They found that this approach increased the PSNR by more than 4 dB. Another obvious solution is to use faster hardware such as GPU. This option has been explored in many studies. For instance, implementation of an NLM-type algorithm on GPU reduced the computational time by a factor of 35 in one study \cite{li2014b}. Iterative reconstruction algorithms with non-local patch-based regularization terms have also been implemented on GPU \cite{tian2011, jia2012}. Another remarkable example was shown in \cite{bartuschat2009}, where the authors implemented the K-SVD algorithm for CT denoising on Cell Broadband Engine Architecture and achieved speedup factors between 16 and 225 compared with implementation on CPU.

There have also been many algorithmic approaches to reducing the computational time. An ingenious and highly efficient method to address this challenge was proposed in \cite{rubinstein2010b}. This method, which is named ``double sparsity" is based on the observation that the learned dictionary atoms, themselves, have a sparse representation in a standard basis, such as DCT. The authors suggest a dictionary structure of the form $D= \Phi A$, where $\Phi$ is a basis with fast implicit implementation and $A$ is a sparse matrix. They show that this dictionary can be efficiently learned using an algorithm similar to the K-SVD algorithm. Denoising of 3D CT images with this dictionary structure leads to speed-up factors of around 30, while also improving the denoising performance. A relatively similar idea is the separable dictionary proposed in \cite{hawe2013}, where the dictionary to be learned from data is assumed to be the Kronecker product of two smaller dictionaries. By reducing the complexity of sparse coding from $\mathcal{O}(n)$ to $\mathcal{O}(\sqrt{n})$, this dictionary model allows much larger patch/block sizes to be used, or alternatively, it results in significant speedups for equal patch size. A two-level dictionary structure was proposed in \cite{li2012}. In this method, the learned dictionary atoms are clustered using a k-means algorithm using the coherence as the distance measure. For sparse coding of a test patch, a greedy algorithm is used to select the most likely atoms which are then used to obtain the sparse representation of the patch. Another study used the coherence of the dictionary atoms in learning a dictionary on a graph and reported very good results in 3D CT denoising \cite{li2012b}.

For dictionary-based methods, the most computationally demanding part of the algorithm during both dictionary learning and usage is the sparse coding step. As we mentioned above, the image is usually divided into overlapping patches/blocks and the sparse representation of each patch/block in the dictionary has to be computed at least once (more than once if the algorithm is iterative). If the dictionary has no structure, which is the general case for overcomplete learned dictionaries, the sparse coding of each patch will require solving a small optimization problem. This will be computationally demanding, especially when the number and size of these patches/blocks is large such as in 3D CT. In recent years, many algorithms have been suggested for sparse coding of large signals in unstructured dictionaries. Some of these algorithms are basically faster implementations of traditional sparse coding algorithms \cite{krstulovic2006, rubinstein2008}, while others are based on more novel ideas \cite{lee2006, gregor2010, xiang2011, bronstein2012}. Some of these methods have achieved several orders of magnitude speedups \cite{gregor2010, bronstein2012}. A description of these algorithms is beyond the scope of this manuscript, but the computational edge that they offer makes patch-based methods more appealing for large-scale CT imaging.

For the NLM algorithms, the major computational bottleneck is the search for similar patches. We described some of the state-of-the-art methods for reducing the computational load of patch search in Section \ref{sec:NLM}. There has been little published research on how these techniques may work on CT images. One study applied the method of integral image \cite{darbon2005b} on CT images. The same study reported that if the smoothing strength is properly adjusted, a very small search window and a very small patch size can be used, leading to significant savings in computation.

\section{Conclusions}
\label{sec:conclusions}

Sparsity-based models have long been used in digital image processing. Recently, learned overcomplete dictionaries have been shown to lead to better results than analytical dictionaries such as wavelets in almost all image processing tasks. Nonlocal patch similarities have also been proven to be extremely useful in many image processing applications. Algorithms based on nonlocal patch similarities are considered to be the state of the art in important applications such as denoising. The practical utility of patch-based models has been demonstrated by hundreds of studies in recent years, many of which have been conducted on medical images. Use of learned overcomplete dictionaries for sparse representation of image patches and use of nonlocal patch similarities are at the core of much of the ongoing research in the field of image processing.

The published studies on the application of these methods for reconstruction and processing of CT images have reported very good results. However, the amount of research on the application of these methods in CT has been far less than that on natural images. Any reader who is familiar with the challenges of reconstruction and processing of CT images will acknowledge that there is an immense potential for these methods to improve the current state of the art algorithms in CT.

In terms of the pre-processing algorithms, there has been only a couple of published papers on patch-based algorithms. This is partly due to the fact that most of the patch-based models and algorithms have been originally proposed for uniform Gaussian noise. For instance, greedy sparse coding algorithms that are a central component of methods that use learned overcomplete dictionaries have been proposed for the case of Gaussian noise. As we mentioned in Section \ref{sec:poisson-patch}, only recently similar methods for the case of Poisson noise have started to appear. Nonetheless, even with the current tools, patch-based models can serve as very useful tools for developing powerful pre-processing algorithms for CT. Some of the patch-based methods that we have reviewed in Section \ref{sec:poisson-patch} have been applied on very noisy images (i.e., very low-count Poisson noise) and they have achieved impressive results. This might be extremely useful for low-dose CT that is of especial interest in clinical settings.

Iterative CT reconstruction algorithms that have used TV or patch-based regularization terms have reported very promising results. One can say that the published works have already demonstrated the usefulness of patch-based methods for CT reconstruction. However, many of the proposed algorithms have been applied on 2D images and in some cases it is not clear if the proposed algorithm can be applied to large 3D reconstruction where efficient implementations of forward and back-projection operations limit the type of iterative algorithm that can be employed. Moreover, little is known about the robustness of these algorithms in terms of the trained dictionary. As we mentioned in Section \ref{sec:DL_learning}, the dictionary learning problem is non-convex and, hence, the dictionary learning algorithms are not supported by strong theoretical guarantees.

Post-processing accounts for the largest share of the published papers on the application of patch-based methods in CT. Both denoising and restoration (e.g., artifact removal) algorithms have been proposed. Even though most of these papers have reported good results, many of them have used algorithms that have been originally proposed for natural images with little modification. Therefore, it is likely that much better results could be achieved by designing dedicated algorithms for CT. In fact, CT images, especially those reconstructed from low-dose scans, present unique challenges. Specifically, these images are contaminated by very strong noise with a non-uniform and unknown distribution. Moreover, they are also marred by various types of artifacts. This situation calls for carefully-devised algorithms that are tailored for CT images. Although this can be challenging, the success of patch-based methods on natural images can be taken as a strong indication of their potential to tackle these challenges. Patch-based methods have led to the best available denoising algorithms. Moreover, they have been successfully used for suppressing various types of artifacts and anomalies in natural images and videos. Therefore, they are likely to achieve state of the art denoising and restoration results in CT.

\bibliographystyle{IEEEtran}
\bibliography{davoodreferences}

\end{document}